\documentclass[12pt]{article}

\usepackage{graphics}
\usepackage{graphicx}
\usepackage{longtable}
\usepackage{amssymb}
\usepackage{amsmath}
\usepackage{subfigure}

\def\r{\textbf{r}}

\def\d{\textrm{d}}
\def\e{\textrm{e}}

\title{Complexity and neutron stars structure}

\author{K.Ch. Chatzisavvas, V.P. Psonis, C.P. Panos, Ch.C.
Moustakidis
\\
 {\it  Department of Theoretical Physics,}\\
        {\it Aristotle University of Thessaloniki,}\\
                {\it  54124 Thessaloniki, Greece}
 }

\date{June, 2008}

\begin{document}

\maketitle

\begin{abstract}
We apply  the statistical measure of complexity introduced by
L\'{o}pez-Ruiz, Mancini and Calbet \cite{Lopez95} to neutron stars
structure. Neutron stars is a classical example where the
gravitational field and quantum behavior are combined and produce
a macroscopic dense  object. Actually, we continue the recent
application of  Sa\~{n}udo and Pacheco \cite{Sanudo09} to white
dwarfs structure. We concentrate our study on the connection
between complexity and neutron star properties, like maximum mass
and the corresponding radius, applying a specific set of realistic
equation of states. Moreover, the effect of the strength of the
gravitational field on the neutron star structure and consequently
on the complexity measure is also investigated. It is seen that
neutron stars, consistent with astronomical observations so far,
are ordered systems (low complexity), which cannot grow in
complexity as their mass increases. This is a result of the
interplay of gravity, the short-range nuclear force and the very
short-range weak interaction.
\end{abstract}


\section{Introduction}

Information theory, founded by Shannon to provide a theoretical
framework in communications \cite{Shannon48}, has been further
employed as a useful tool to characterize physical systems during
the next decades
\cite{Bialynicki75,Garde85,Garde87,Ghosh84,Ohya93}. Since then, a
series of studies concerning the application of information theory
to various physical systems has appeared in the literature, while
their number exhibits a remarkable rise over the past decade. The
dependence of various information-theoretic measures on some
parameters of the physical systems has been studied, the presence
of correlations has been quantified, connections with experimental
data have been detected and universal properties have been
extracted. Very recently, these investigations have been extended
to include statistical complexity measures, in order to examine
self-organizing characteristics of physical systems, patterns and
correlations. Although a complete and universal definition for
complexity is missing, the current framework provides interesting
and  satisfactory results expected from intuition. So far, various
complexity measures are used taking into account conditions and
constraints imposed by the physical system under consideration
\cite{Anteneodo97,Crutch98,Crutch00,Binder00}. Such information
and complexity studies (focusing so far on the two statistical
complexity measures SDL and LMC
\cite{Lopez95,Landsberg-98,Shiner-99}) have been applied to
various quantum many-body systems i.e. nuclei, atoms, atomic
clusters,  bosons and molecules
\cite{Chatzisavvas05}-\cite{Sagar08}. Recently San\~{u}do and
Pacheco \cite{Sanudo09} were the first to extend those studies to
an astronomical object i.e. a white dwarf. Specifically, the
Shannon information entropy $S$ and the statistical complexity $C$
have been calculated in the two kinematics extremes
(non-relativistic and relativistic cases) of the
electron-degenerate matter of a dwarf.

In the present work we study the information content of another
astronomical object, a neutron star. With a mass of 1.4 up to 3
solar masses, a radius of $\sim10$ Km and an average density of
$10^{14}$ g/cm$^3$, the neutron star is one of the possible
endpoints of stellar evolution. Further gravitational collapse is
counterbalanced by repulsive forces originating from Pauli's
exclusion principle, if the mass of the compressed stellar core is
less than the Oppenheimer-Volkoff limit of about 3 solar masses.

Neutron stars are systems with several similarities with atomic
systems, but there are also fundamental differences. In an atomic
system self-organization is reached through the competition
between the Coulomb interaction and the Pauli principle. In fact,
the long-range electromagnetic interaction is the main interaction
among the particles of the system. In addition, atoms are
microscopic systems with a typical dimension of a few Angstroms
($10^{-10} \,$m). In contrast, a neutron star is a macroscopic
system with typical dimension of $10^4 \,$m, much more complicated
than the atoms, in the sense that it is organized under the
competition of mainly the following three interactions. The
long-range force of gravity, whose pressure tends to compress the
mass of the star. The short-range nuclear interaction, which
through the degeneracy pressure of the nucleons tends to extend
the outer mass of the star. Finally, the very short-range weak
interaction, which is a kind of regulator of the particle fraction
and thus affects indirectly the properties of the neutron star.
The above forces coexist in harmony in the interior of a neutron
star. The main features of the structure of a neutron star (mass,
pressure and radius) are described by the
Tolman-Oppenheimer-Volkoff equations \cite{Tolman-39}, while they
also depend strongly  on the applied nuclear equation of state.

In this Letter we present a study of the information properties of
a neutron star and  explore how they are connected with the
characteristic properties of the structure of the system, i.e. its
mass $M$ and radius $R$. Furthermore, we investigate the
dependence of information and complexity measures on the nuclear
forces, through the asymmetry energy parameter $c$, and  the
gravitational constant $G$. Also we comment on the effect of
information measures on the stability of a neutron star, based on
the fact that stability regions are characterized by the
inequality $\d M/ \d R <0$.

Here, we consider that the temperature of a star is $T=0$, in the
context that the Fermi energy is much greater than $kT$. However,
it should be of interest to extend our study and try to connect
the thermodynamic properties of a hot neutron start with the
information content of the system. Furthermore, it is important to
examine the information properties of other astronomical objects
e.g. stars consisting of fermions or bosons, with arbitrary masses
and interaction strengths. Such a work is in progress.

The outline of this Letter is the following: In Section
\ref{sec:2}, we define the information and complexity measures
employed here, together with a model of neutron stars. In Section
\ref{sec:3}, we present our results and a discussion, while
Section \ref{sec:4} contains a summary.

\section{The model}\label{sec:2}
\subsection{Theoretical information measures}

The Shannon information entropy $S$ \cite{Shannon48} for a
continuous probability distribution $\rho(\r)$, denoting a measure
of the amount of uncertainty associated with a probability
distribution, is defined as
\begin{equation}
    S=-\int \rho(\r) \, \ln{\rho(\r)} \, \d \r, \label{S-1}
\end{equation}
while the disequilibrium $D$, being a quadratic distance from
equiprobability, is
\begin{equation}
    D=\int \rho^2(\r) \, \d \r , \label{D-1}
\end{equation}
with dimension of inverse volume.

For a continuous probability distribution the disequilibrium is
indeed the same measure as the \emph{information energy} defined
by Onicescu \cite{Onicescu66}.

For discrete probability distributions
$\{p_i\}=\{p_1,p_2,\cdots,p_N\}$, the information entropy
$S=-\sum_{i=1}^{N} p_i \,\ln{p_i}$ is minimum $(S_{\rm min}=0$)
for the distribution of a completely regular system (absolutely
localized), where one of the $p_i$'s equals unity, while all the
others vanish. The maximum value $(S_{\max} =\ln{N})$ is attained
for the equiprobable distribution (completely delocalized), where
$p_i =1/N, \, i=1,\ldots,N$. On the other hand, the disequilibrium
$D=\sum_{i=1}^{N} (1/N -p_i)^2$, is maximum, $D_{\rm max}=1-1/N
\rightarrow 1$ (for large $N$) for a completely regular system,
while it is minimum, $D_{\rm min}=0$ for an equiprobable
distribution.

In the continuous case, an equiprobable probability distribution
can be defined as a rectangular function, while a completely
regular system corresponds to a $\delta$-like probability
distribution function, where the width of the distribution becomes
very narrow and its peak extremely high.

In order to study the statistical complexity defined by
L\'{o}pez-Ruiz, Mancini and Calbet (LMC) \cite{Lopez95}, we use a
slightly modified definition introduced in \cite{Catalan02}
\begin{equation}
    C=H\cdot D, \label{C-1}
\end{equation}
where
\begin{equation}
    H=\e^{S}, \label{H-1}
\end{equation}
is the information content of the system, while the exponential
functional preserves the positivity of $C$.

The aforementioned definitions of information entropy and
disequilibrium in the case of neutron stars  are modified as
follows:

\begin{equation}
    S=-b_0\int \bar{\epsilon}(r) \, \ln \bar{\epsilon}(r) \, \d\r,
    \label{S-2}
\end{equation}
and
\begin{equation}
    D=b_0 \int \bar{\epsilon}(r)^2 \, \d\r , \label{D-2}
\end{equation}
where $b_{0}=8.9 \times 10^{-7}$ Km$^{-3}$ is a proper constant
satisfying the condition that both information entropy $S$ and
disequilibrium should be dimensionless quantities, while
$\bar{\epsilon}(r)$ is the dimensionless energy density of the
system. It is equivalent to the density mass $\rho(r)$, obtained
by solving the structure equations characterizing the system.

\subsection{Neutron star structure equations}
In order to calculate the gross properties of a neutron star, we
assume that the star has a spherically symmetric distribution of
mass in hydrostatic equilibrium and is extremely cold ($T=0$).
Effects of rotations and magnetic fields are neglected and the
equilibrium configurations are obtained by solving the
Tolman-Oppenheimer-Volkoff (TOV) equations \cite{Tolman-39}
\begin{eqnarray}
    \frac{dP(r)}{dr}&=&-\frac{G M(r) \rho(r)}{r^2}
    \left(1+\frac{P(r)}{c^2 \rho(r)}\right) \left(1+\frac{4 \pi r^3
    P(r)}{c^2 M(r)} \right) \left(1-\frac{2 G M(r)}{c^2 r}
    \right)^{-1} , \nonumber \\
    \nonumber \\
    \frac{d M(r)}{dr}&=&4 \pi r^2 \rho(r)=\frac{4 \pi r^2
    \epsilon(r)}{c^2},
    \label{TOV-1}
\end{eqnarray}
where $P(r)$ and $M(r)$ are the pressure and the mass functions of
the star respectively.

To solve the set of equations (\ref{TOV-1}) for $P(r)$ and $M(r)$,
one can integrate outwards from the origin ($r=0$) to the point
$r=R$, where the pressure  becomes zero. This point defines R as
the radius of the star. To do this, one needs an initial value of
the pressure at $r=0$, called $P_c=P(r=0)$. The radius $R$ and the
total mass of the star, $M\equiv M(R)$, depend on the value of
$P_c$. To be able to perform the integration, one also needs to
know the energy density $\epsilon(r)$ (or the density mass
$\rho(r)$) in terms of the pressure $P(r)$. This relationship is
the equation of state (EOS) for neutron star matter and here, has
been calculated applying a phenomenological nuclear model.

We can modify equations (\ref{TOV-1}), so that they become
suitable for programming, in the following form:
\begin{eqnarray}
    \frac{\rm{d}\bar{P}(r)}{\rm{d}r}&=&-1.474 \, \frac{\bar{\epsilon}(r)
    \bar{M}(r)}{r^2}
    \left(1+\frac{\bar{P}(r)}{\bar{\epsilon}(r)}\right) \left(1+11.2
    \times 10^{-6}\, r^3 \frac{\bar{P}(r)}{\bar{M}(r)} \right) \nonumber
\\
    &\times& \left(1-2.948 \, \frac{\bar{M}(r)}{r}
    \right)^{-1} , \nonumber \\
    \nonumber \\
    \frac{\rm{d} \bar{M}(r)}{\rm{d}r}&=&11.2\times 10^{-6}\  r^2 \ \bar{\epsilon}(r).
    \label{TOV-2}
\end{eqnarray}
In Eqs.~(\ref{TOV-2}), the quantities $\bar{P}(r)$,
$\bar{\epsilon}(r)$ and $\bar{M}(r)$ are dimensionless. The radius
$r$ is measured in Km. More specifically:
\begin{equation}
    M(r)=\bar{M}(r)\,M_{\odot}, \quad
    \epsilon(r)=\bar{\epsilon}(r)\, \epsilon_0,
    \quad  P(r)=\bar{P}(r) \, \epsilon_0,
    \quad \epsilon_0=1 \, \rm{MeV} \, \rm{fm}^{-3}.
    \label{quant-1}
\end{equation}
It is obvious from Eqs.~(\ref{TOV-2}) that
\begin{equation}
    \bar{M}(R)=11.2\times 10^{-6} \, \int_0^R   r^2 \, \bar{\epsilon}(r)
    \d r = b_0\int \bar{\epsilon}(r) \, \d\r . \label{M-1}
\end{equation}
%
%

%
%
%

\subsection{Nuclear equation of state}
In general, the energy per baryon of neutron-rich matter may be
written to a very good approximation as
\begin{equation}
    \frac{E(n,x)}{A}=\frac{E(n,\frac{1}{2})}{A}+(1-2x)^2 E_{\rm sym}(n) \
    , \label{En-1}
\end{equation}
where $n$ is  the baryon density ($n=n_n+n_p$) and  $x$ is the
proton fraction ($x=n_p/n$). The symmetry energy $E_{\rm sym}(n)$
can be expressed in terms of the difference of the energy per
baryon between neutron ($x=0$) and symmetrical ($x=1/2$) matter.
Here, we consider a schematic equation for symmetric nuclear
matter energy (energy per baryon $E/A$ or equivalently the energy
density per nuclear density $\epsilon/n$), given by the expression
\cite{Prakash-88}
\begin{equation}
    \frac{E(n,1/2)}{A}=\frac{\epsilon_{\rm sym}}{n}=m_Nc^2
    +\frac{3}{5}E_F^0 u^{2/3}+V(u), \quad u=n/n_0 , \label{ESNM-1}
\end{equation}
where $E_F^0=(3/5) (\hbar k_F^0)^2/(2m_N)$ is the mean kinetic
energy per baryon in equilibrium state and $n_0$ is the saturation
density.

The density dependent potential $V(u)$ of the symmetric nuclear
matter is parameterized, based on the previous work of Prakash et.
al. \cite{Prakash-88,Prakash-97}, as follows
\begin{equation}
V(u)=\frac{1}{2} \, Au+\frac{B u^{\sigma}}{1+B' u^{\sigma -1}}+3
\sum_{i=1,2}C_i \left(\frac{\Lambda_i}{p_F^0}\right)^3
\left(\frac{p_F}{\Lambda_i}-\arctan\frac{p_F}{\Lambda_i}\right),
\label{Vu-1}
\end{equation}
where $p_F$ is the Fermi momentum, related to $p_F^0$ by
$p_F=p_F^0u^{1/3}$. The parameters $\Lambda_1$ and $\Lambda_2$
parameterize the finite-range forces between nucleons. The values
employed here are $\Lambda_1=1.5 p_F^0$ and $\Lambda_2=3 p_F^0$.
The parameters $A$, $B$, $B'$, $\sigma$, $C_1$ and $C_2$ are
determined using the constraints provided by the empirical
properties of symmetric nuclear matter at the saturation density
$n_0$. Then, the values of the above parameters are determined in
order that
\begin{equation*}
    E(n=n_0)/A-m_Nc^2=-16 \, \textrm{MeV}, \quad n_0=0.16 \, \textrm{fm}^{-3},
    \quad K_0=240 \, \textrm{MeV}.
\end{equation*}
In general, the parameter values for three possible values of the
compression modulus $K_0$
$\left(K_0=9n_0^2\frac{d^2(E/A)}{dn^2}|_{n_0} \right)$ are
displayed  in Table I, in \cite{Prakash-88}.

To a very good approximation, the nuclear symmetry energy $E_{\rm
sym}$ can be parameterized as follows \cite{Prakash-94}
\begin{equation}
E_{\rm sym}(u) \simeq 13 \, u^{2/3}+17 \, F(u),\label{Esym-3}
\end{equation}
where the first term of the right-hand side part of
Eq.~(\ref{Esym-3}) is  the contribution of the kinetic energy and
the second term comes from the interaction energy. For the
function $F(u)$, that parametrizes the interaction part of the
symmetry energy, we apply the following form
\begin{equation}
    F(u)=u^c ,\label{fu-1}
\end{equation}
where the parameter $c$ (hereafter called potential parameter)
varies between $0.4<c<1.5$ leading to reasonable values for the
symmetry energy. In order to construct the nuclear equation of
state, the expression of the pressure is needed. In general, the
pressure, at temperature $T=0$, is given by the relation
\begin{equation}
    P=n^2 \, \frac{d(\epsilon/n)}{d n}=n \, \frac{d\epsilon}{d n}-\epsilon .
    \label{Pres-1}
\end{equation}
Employing equations (\ref{En-1}), (\ref{ESNM-1}) and
(\ref{Pres-1}), we find the contribution of the baryon to the
total pressure:
\begin{equation}
    P_b=\left[\frac{2}{5}\, E_F^0 \, n_0 \, u^{5/3}+u^2 n_0 \, \frac{d V(u)}{du}
    \right] + n_0\, (1-2x)^2 \, u^2 \, \frac{d E_{\rm sym}(u)}{d u} .
\end{equation}

The leptons (electrons and muons), originating from the condition
of the beta stable matter,  contribute also to the total energy
and total pressure \cite{Prakash-94}. To be more precise, the
electrons and the muons, which are the ingredients of the neutron
star, are considered as non-interacting Fermi gases. In that case
their contribution to the total energy and pressure  is
\begin{align}
     \epsilon_{e^-,\mu^-}&=\frac{m_l^4 c^5}{8 \pi^2
    \hbar^3}\left[(2z^3+z)(1+z^2)^{1/2}-\sinh^{-1}(z)\right],
    \\
     P_{e^-,\mu^-}&=\frac{m_l^4 c^5}{24 \pi^2
    \hbar^3}\left[(2z^3-3z)(1+z^2)^{1/2}+3\sinh^{-1}(z)\right],
    \label{lepton}
\end{align}
where $z=k_F/m_lc$. Now the total energy and total pressure of
charge neutral and chemically equilibrium nuclear matter are
\begin{equation}
\epsilon_{tot}=\epsilon_b+\sum_{l=e^{-},\mu^{-}}\epsilon_l,
\label{tot-e}
\end{equation}

\begin{equation}
P_{tot}=P_b+\sum_{l=e^{-},\mu^{-}} P_l \ .\label{tot-pr}
\end{equation}

From equations (\ref{tot-e}) and (\ref{tot-pr}) we can construct
the equation of state in the form $\epsilon=\epsilon(P)$. In order
to calculate the global properties of the neutron star, i.e. the
radius and mass, we solved numerically the TOV equations
(\ref{TOV-1}) with the given equations of state constructed
employing the present model. For very low densities ($n<0.08 \,$
fm$^{-3}$) we use the equation of state according to Feynman,
Metropolis and Teller \cite{Feynman-49} and also from Baym, Bethe
and Sutherland \cite{Baym-71}.

\section{Results and Discussion}\label{sec:3}

\begin{figure}[h]
\includegraphics[height=7.0cm,width=6.5cm]{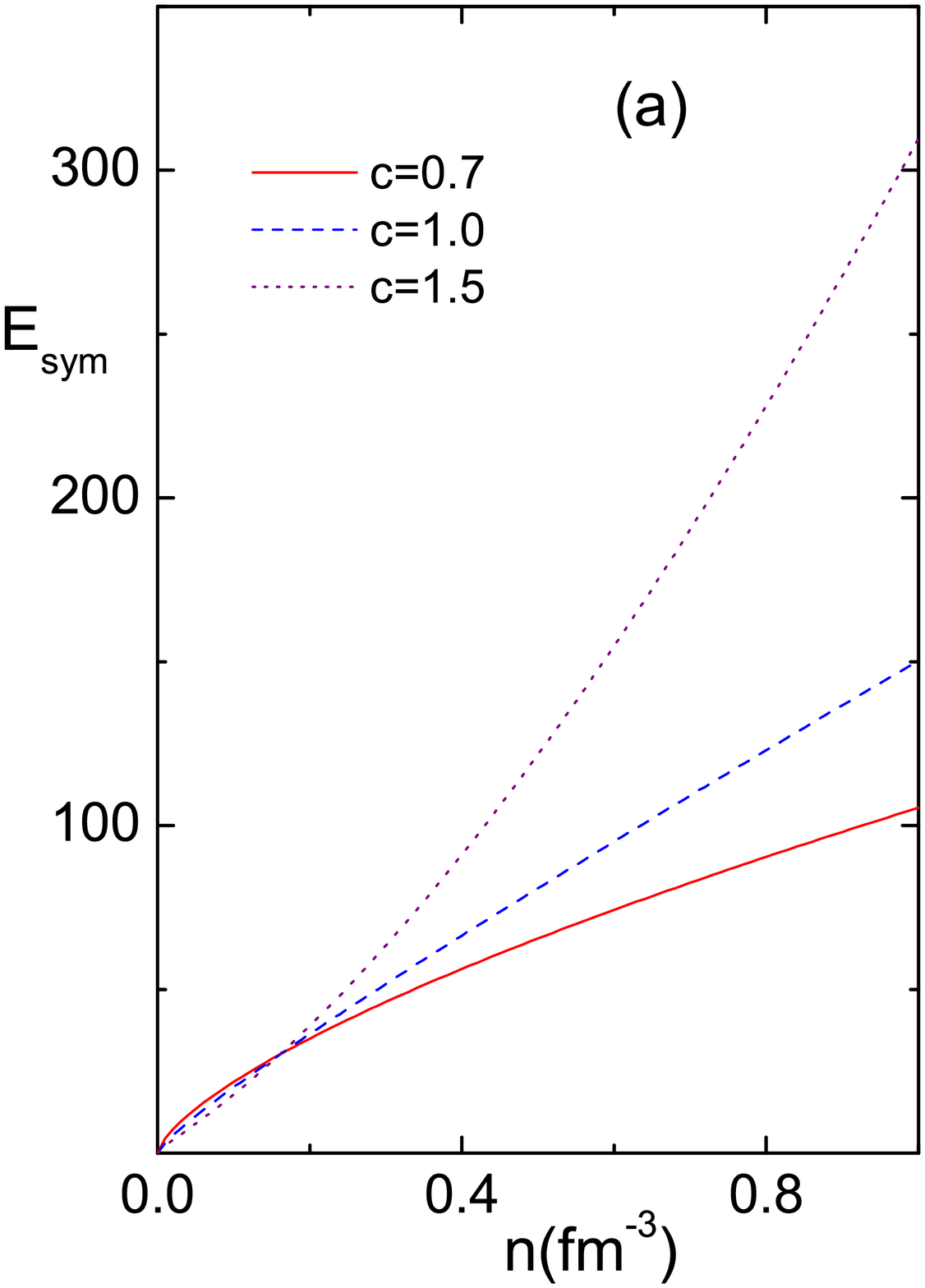}\hspace{-0.8cm}
\includegraphics[height=7.0cm,width=6.5cm]{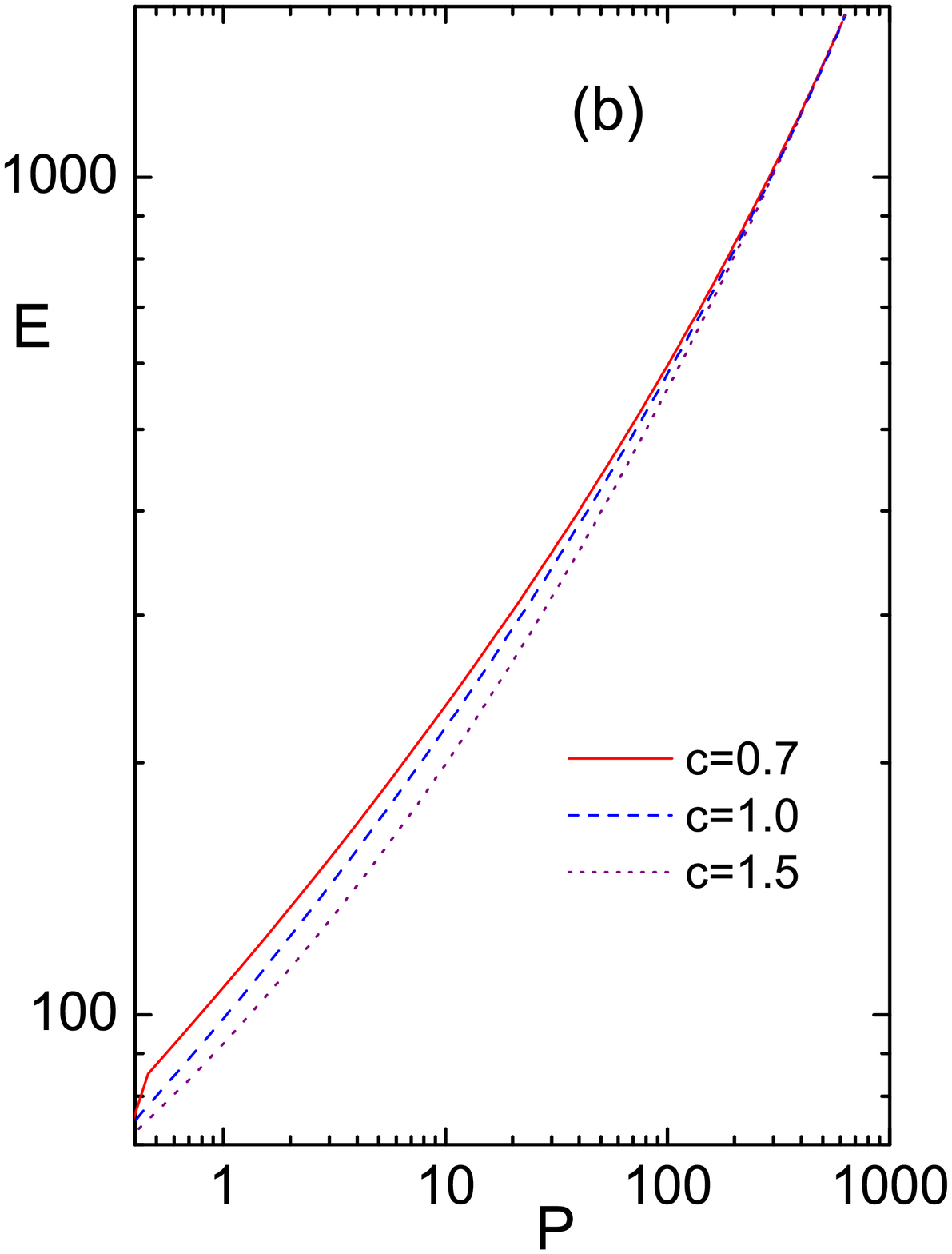}
\\
\vspace{-1.0cm}
\includegraphics[height=7.0cm,width=6.5cm]{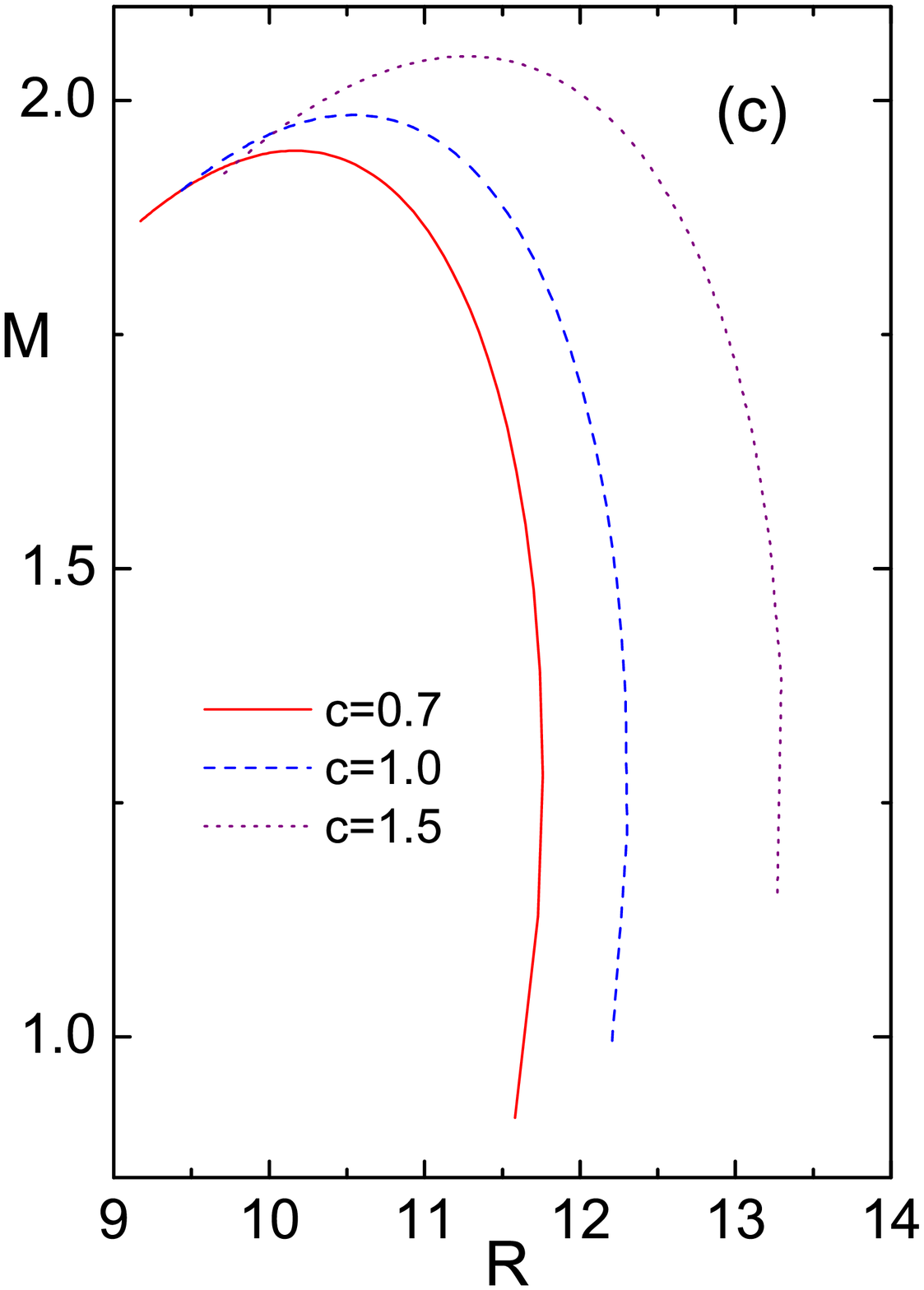}
\caption{(a) Symmetry Energy vs baryon density $n$, (b) Energy vs
Pressure, and (c) Mass vs Radius.} \label{fig:fig1}
\end{figure}

The starting point of our study is the solution of
Eq.~(\ref{TOV-2}) for three different equations of $\beta$-stable
nuclear matter. More precisely, we employ three values of the
parameter $c$, which characterizes the density dependence of the
nuclear symmetry energy, i.e.  $c=0.7$ (soft equation of state),
$c=1.0$, and $c=1.5$ (stiff equation of state). In
Fig.~\ref{fig:fig1}(a), we plot the nuclear symmetry energy
$E_{\rm sym}$, in Fig.~\ref{fig:fig1}(b) the corresponding
equations of state and in Fig.~\ref{fig:fig1}(c) the mass-radius
diagrams for each of the three cases.

\begin{figure}[hb]
\includegraphics[height=7.5cm,width=7.cm]{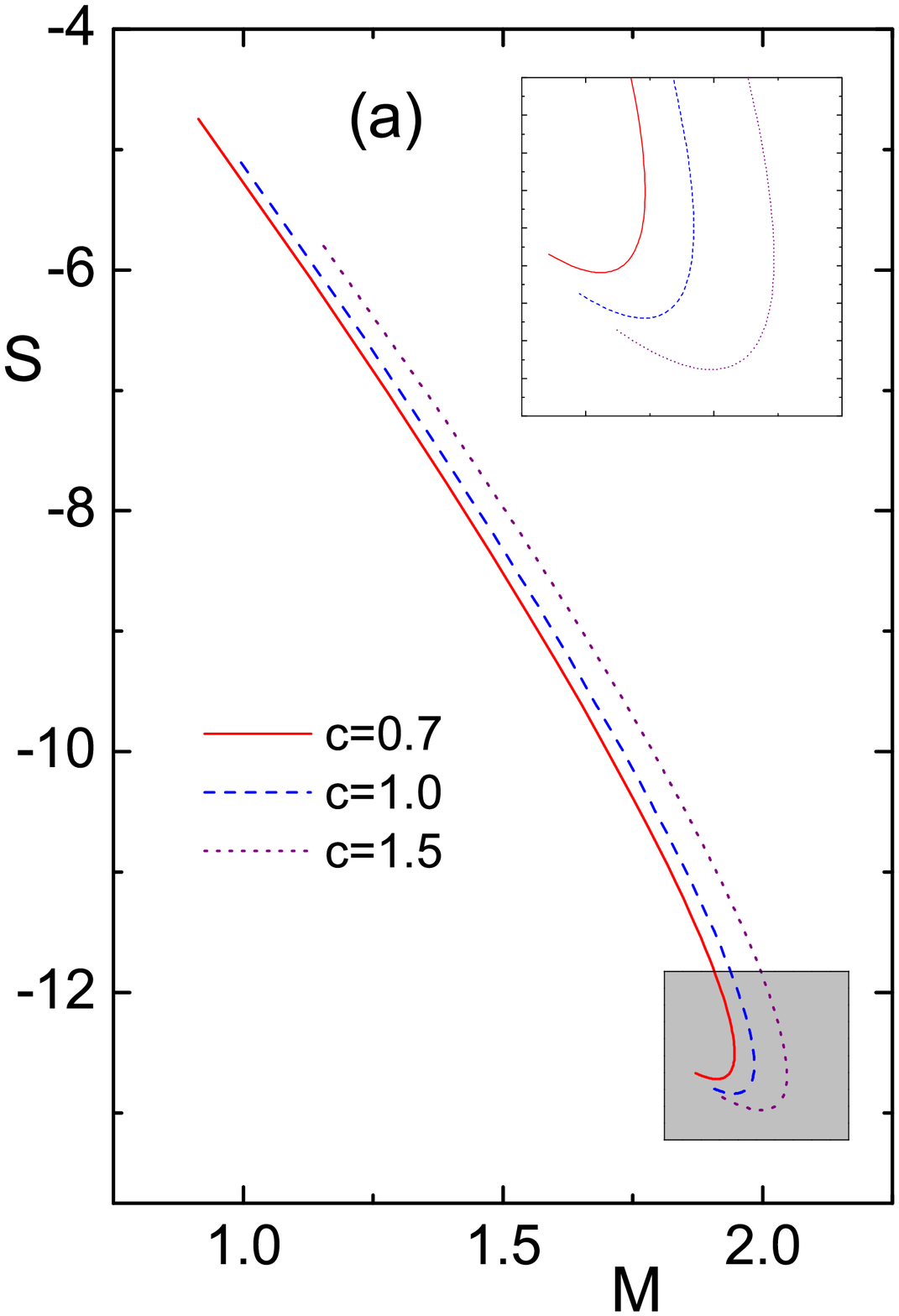}
\includegraphics[height=7.5cm,width=7.cm]{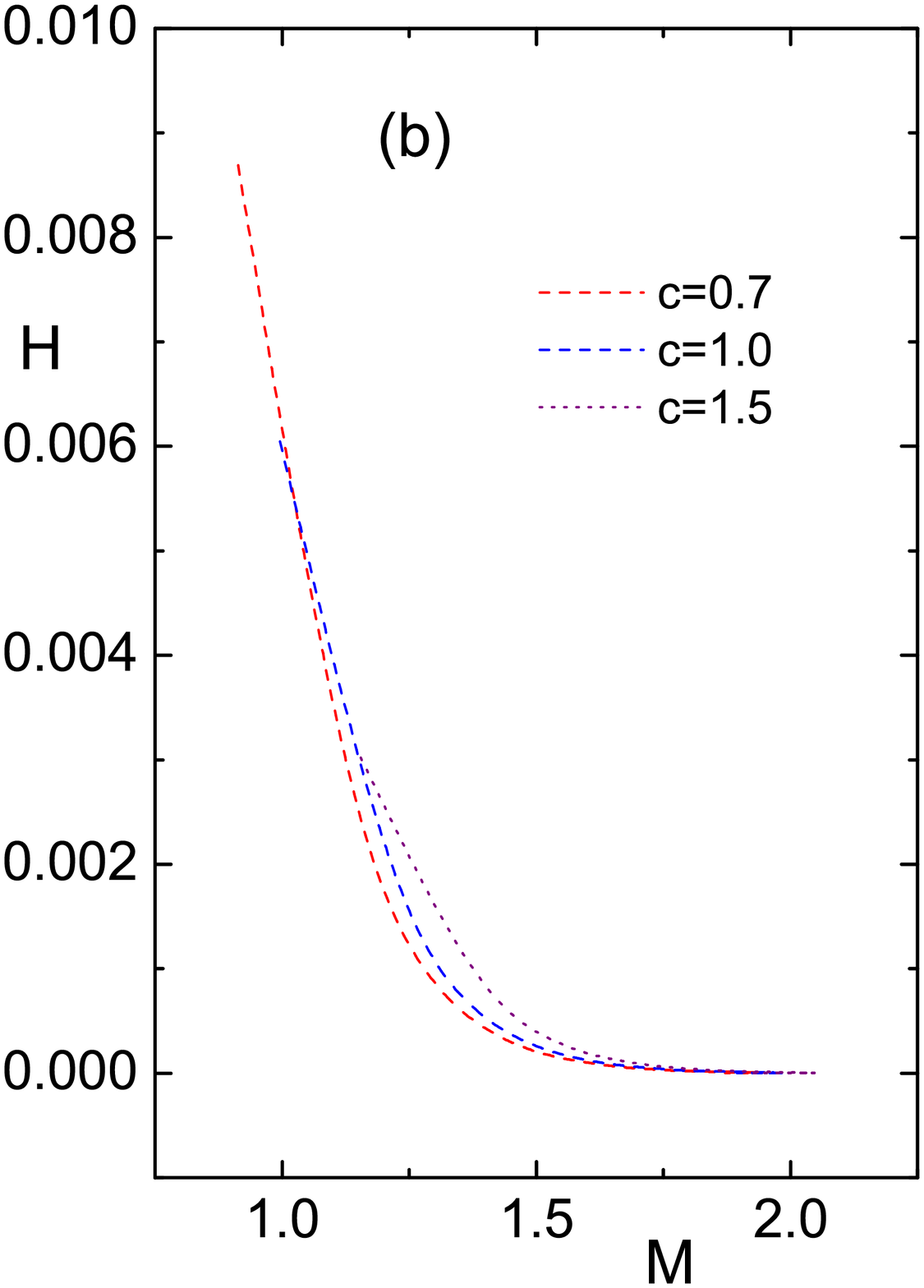}
\\
\includegraphics[height=7.5cm,width=7.cm]{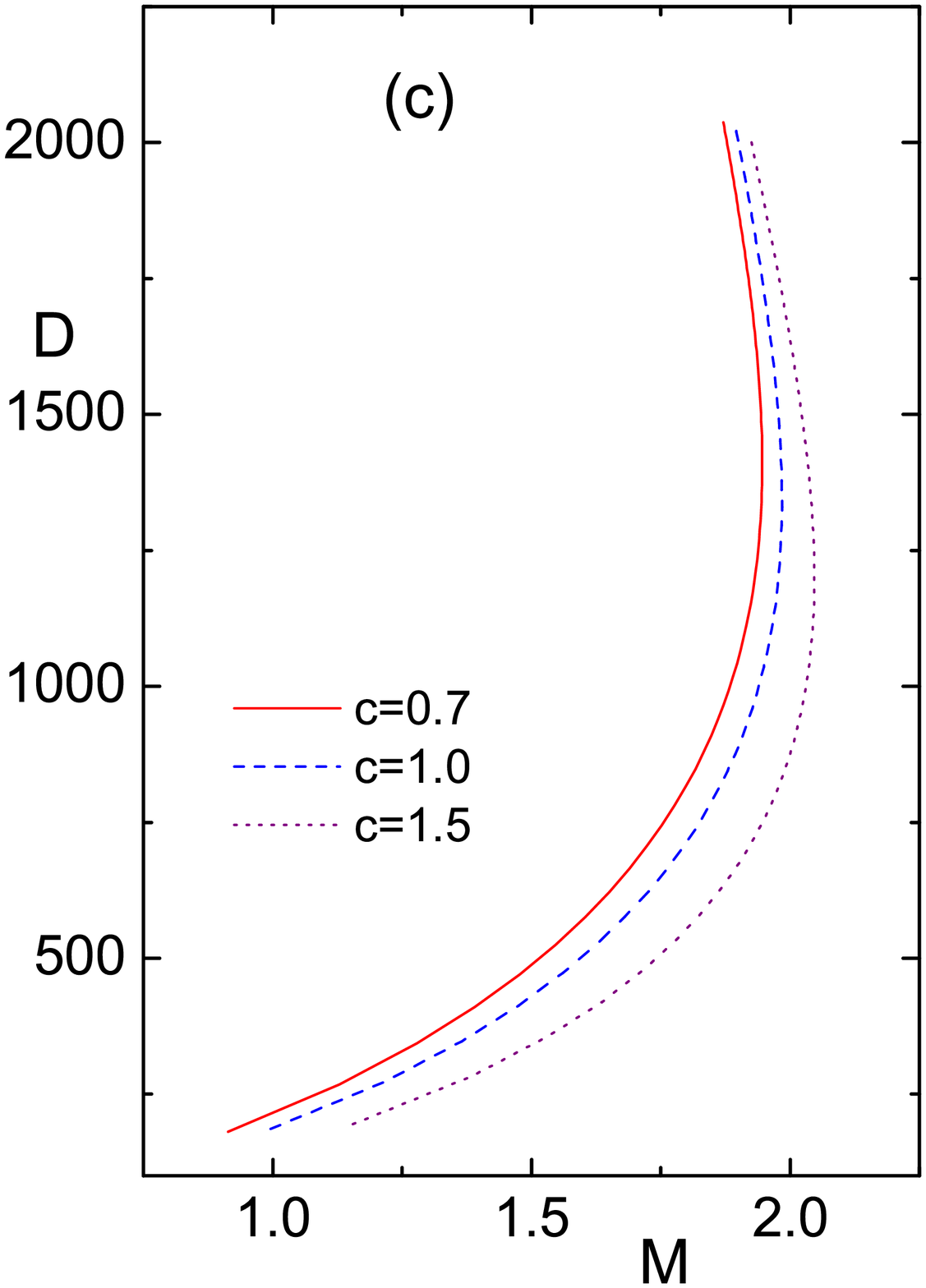}
\includegraphics[height=7.5cm,width=7.cm]{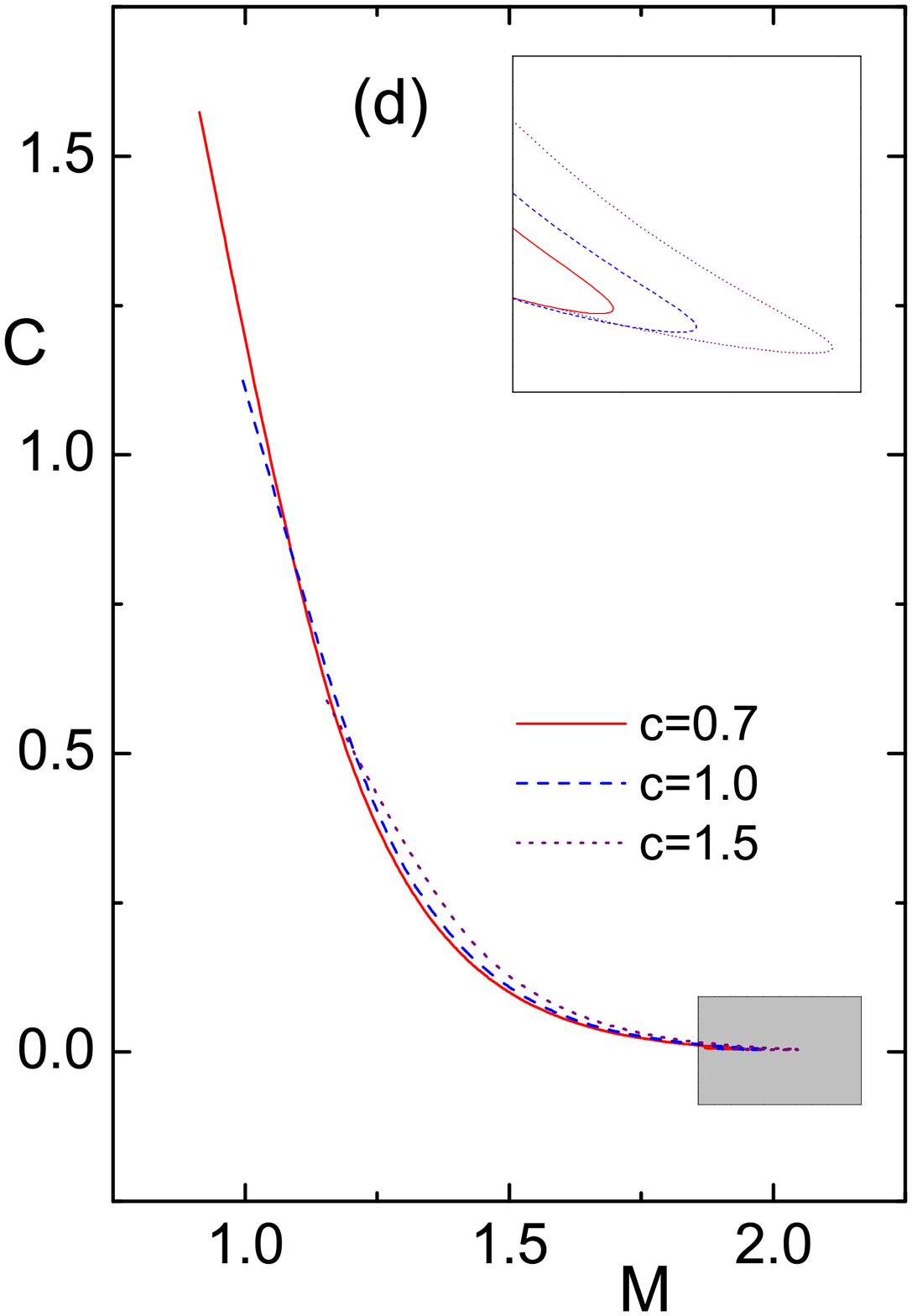}
\caption{(a) Entropy $S(M)$, (b) Information Content $H(M)$, (c)
Disequilibrium $D(M)$, and (d) Complexity $C(M)$. The insets are
commented in the text.} \label{fig:fig2}
\end{figure}

Actually every pair $(R,M)$ in a mass-radius diagram is the
outcome of the structure equations (Eqs.~\ref{TOV-2}) for an
arbitrary chosen initial value of the pressure $P_c$ in the center
of the star. Thus, varying the value of $P_c$ in a reasonable
range, we can have a picture of the behavior of those substantial
structure characteristics. We have to note here that the region
where $dM/dR<0$ corresponds to a stable neutron star, while
$dM/dR>0$ to an unstable one. The presence of the unstable region
($M< M_{\rm max}$ and $dM/dR>0$) seems to lead to double valued
functions $S(M)$ and $C(M)$, for values of $M$ close to $M_{\rm
max}$ (insets in Fig.~\ref{fig:fig2}(a) and
Fig.~\ref{fig:fig2}(d)). However, the study of that region is
beyond the scope of the present work. Another important feature of
a neutron star is the value of the maximum mass $M_{\rm max}$ for
which the star can exist for the specific equation of state. As
displayed in Fig.~\ref{fig:fig1}(c), $M_{\rm max}$ is strongly
dependent on the equation of state, while a stiffer equation leads
to larger $M_{\rm max}$.

In Fig.~\ref{fig:fig2}(a), we present the information entropy $S$,
given by Eq.~(\ref{S-2}), as a function of the mass $M$. We find
that $S$ is a decreasing function of $M$ in the region denoting a
stable neutron star. The above result is a direct consequence of
the fact that when the mass of the star increases, its radius
decreases, so does its volume, while its energy density (or its
mass density) becomes more localized. Thus, the star is less
extended, more compact and $S$ is smaller. The effect of the
parameter $c$ is just to shift the curve $S$ versus $M$.

In Fig.~\ref{fig:fig2}(b) we plot the information content
$H=\e^{S}$ employed in the LMC definition $C=H\cdot D$. It is seen
that both $S(M)$ and $H(M)$ exhibit the same monotonic trend as
functions of $M$.

In Fig.~\ref{fig:fig2}(c) we display the disequilibrium $D(M)$.
Increasing $M$ corresponds to a more concentrated density
distribution, its energy density becomes more localized, resulting
to a monotonically increasing $D$. The rate of this increase is
clearly greater in the region close to the value of $M_{\rm max}$.
This is due to the fact that as $M$ approaches to its maximum
value, it becomes almost independent of $R$. We also observe a
reciprocal behavior of the trends of $S(M)$ and $D(M)$, as
expected from their definitions.

Complexity $C$, is plotted in Fig.~\ref{fig:fig2}(d). In the
region denoting a stable neutron star, $C$ is a monotonically
decreasing function of the star mass $M$. The most interesting
result in this figure is that a neutron star can not grow in
complexity as its mass increases towards the limit of $M_{\rm
max}$. Considering the fact that in nature the most probable
values for the mass of a neutron star vary between 1.4 M$_{\odot}$
and 3 M$_{\odot}$, we note that a neutron star is eventually a
physical system of minimum complexity. It is an ordered system,
since in the corresponding region the rate of decrease of $C$
becomes very small and can be considered as a plateau of minimum
(zero) complexity.

\begin{figure}[h]
\includegraphics[height=10.cm,width=7.cm]{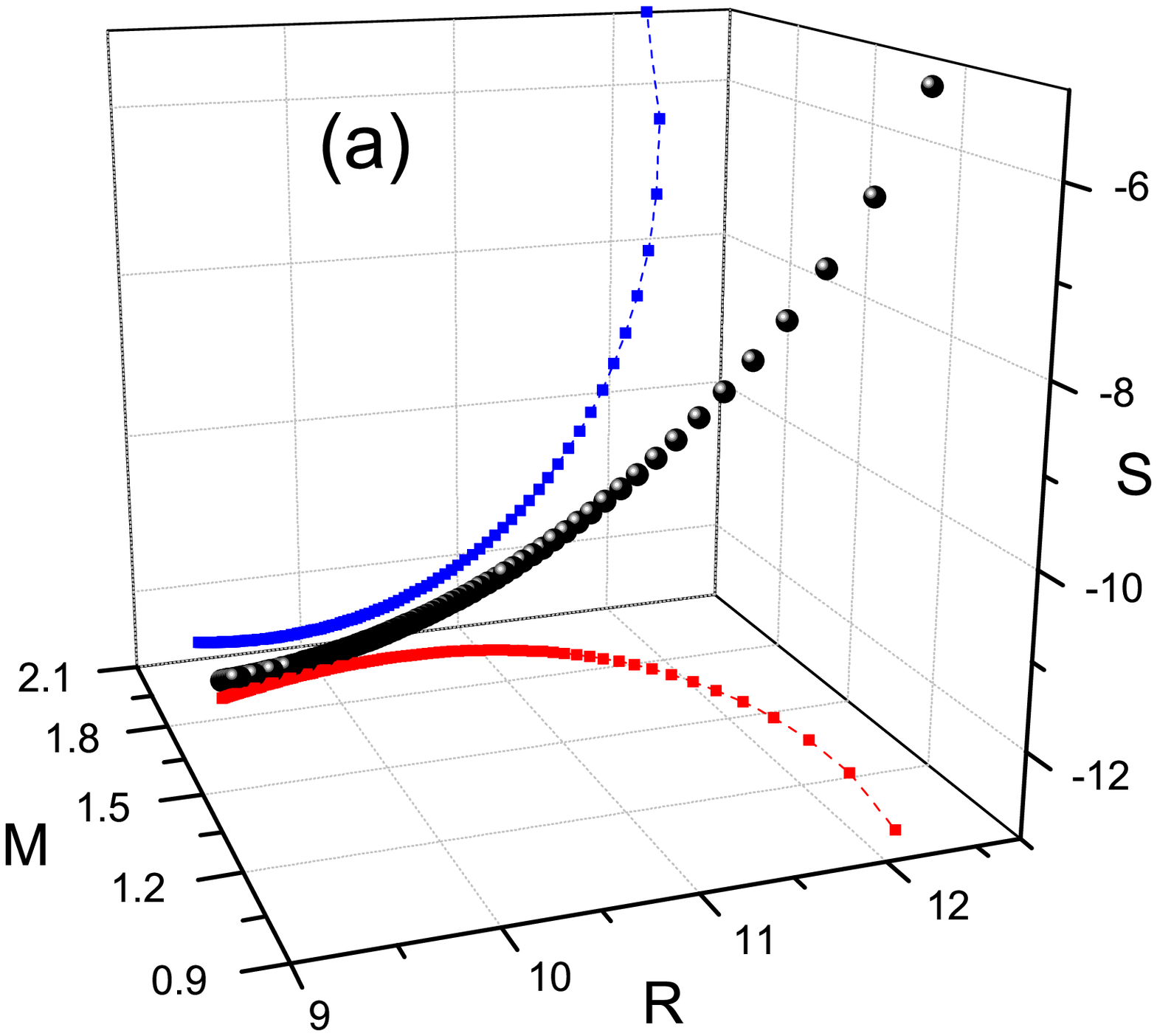}\hspace{-0.8cm}
\includegraphics[height=10.cm,width=7.cm]{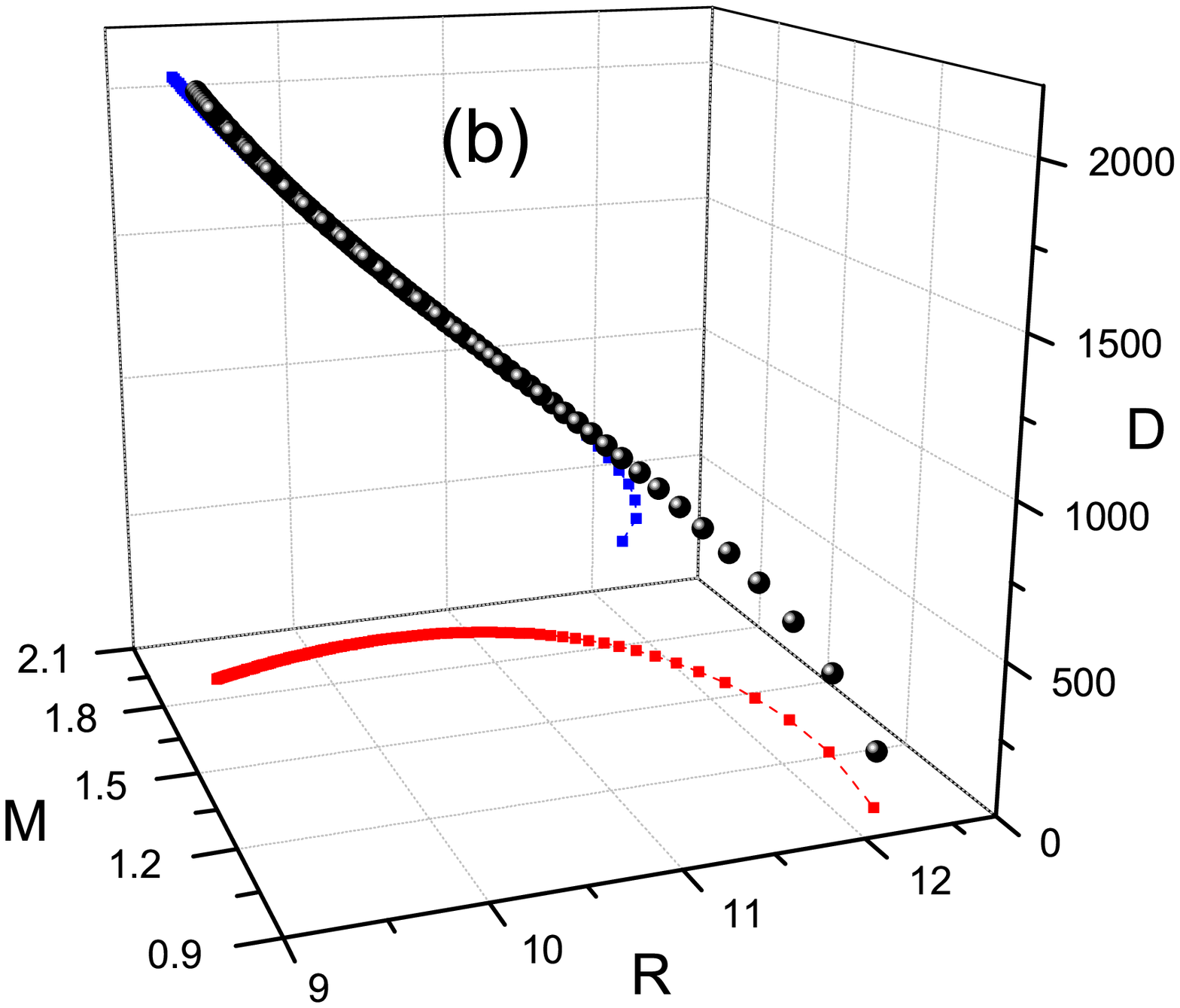}\hspace{-0.8cm}
\includegraphics[height=10.cm,width=7.cm]{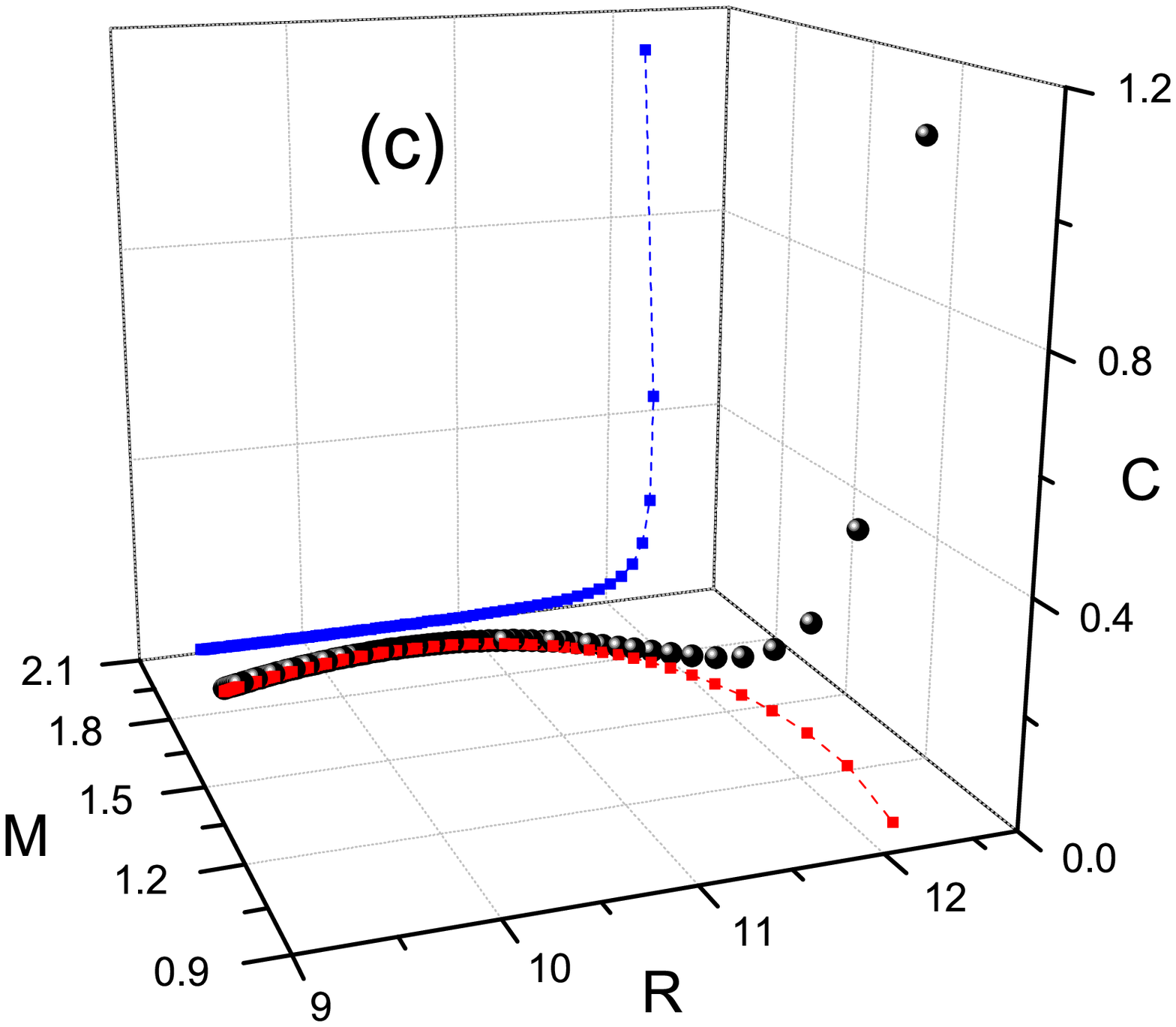}
\caption{3D display of (a) Entropy $S(R,M)$, (b) Disequilibrium
$D(R,M)$, and (c) Complexity $C(R,M)$, projected for each case on
two planes: (a) $R-M$ and $S-R$, (b) $R-M$ and $D-R$, (c) $R-M$
and $C-R$.} \label{fig:fig3}
\end{figure}

This result becomes more striking in the following set of figures,
Fig.~\ref{fig:fig3}, where we plot in three-dimensions (3D)
information and complexity measures, as functions of both $M$ and
$R$, taking advantage of the fact that each choice of initial
values in the equation of state provides a different pair ($R,M$),
reflecting the competition  between  the gravitational and
degenerate gas pressures. The facts that the most probable radii
of a neutron star are close to 10 \,Km, together with the
aforementioned comment on the most likely masses, lead us to
conclude that a neutron star is in general, a system of minimum
complexity. Furthermore, it can not grow in complexity as the mass
or radius increase inside the regions imposed and commented above.
The neutron star is an ordered system. From the 3D plots  of
Fig.~\ref{fig:fig3} we can visualize the variation of $S$ and $D$
as functions of $R$, by keeping $M$ constant. Information entropy
$S$ is an increasing function of $R$, i.e. a larger radius
corresponds to a larger volume, the energy density (or its mass
density) becomes more delocalized, and hence the system is more
extended and the information describing it increases. On the other
hand, the disequilibrium $D$ is a monotonically decreasing
function of $R$ corroborating the fact that the system tends to
equiprobability as $R$ increases.

\begin{figure}[h]
\includegraphics[height=7.5cm,width=7.cm]{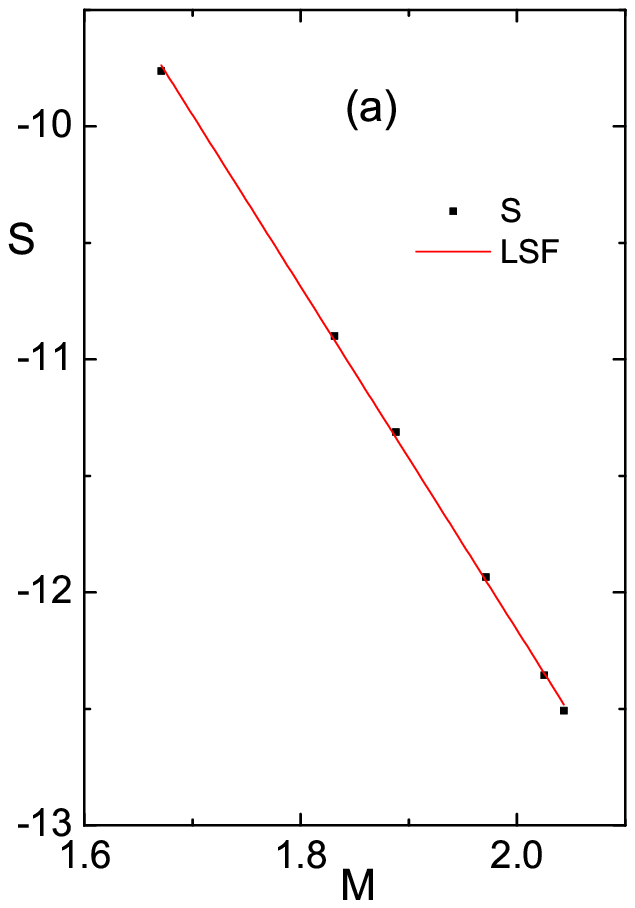}
\includegraphics[height=7.5cm,width=7.cm]{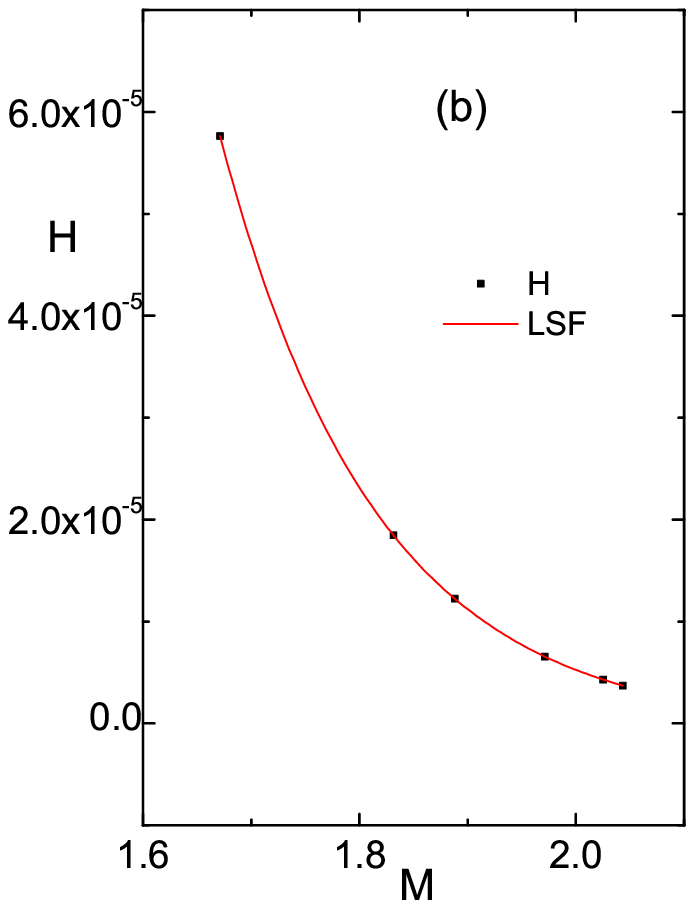}
\\
\includegraphics[height=7.5cm,width=7.cm]{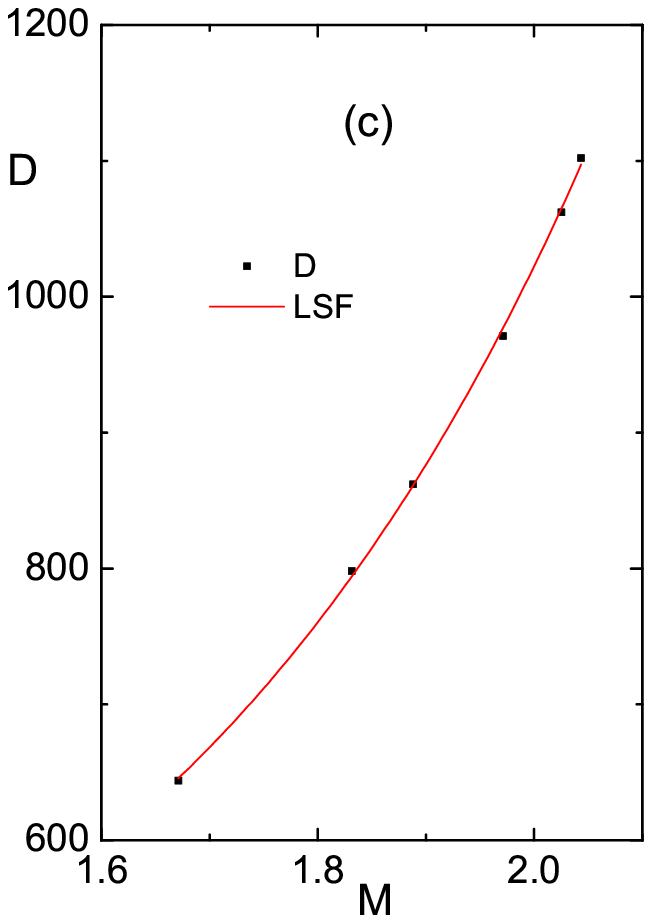}
\includegraphics[height=7.5cm,width=7.cm]{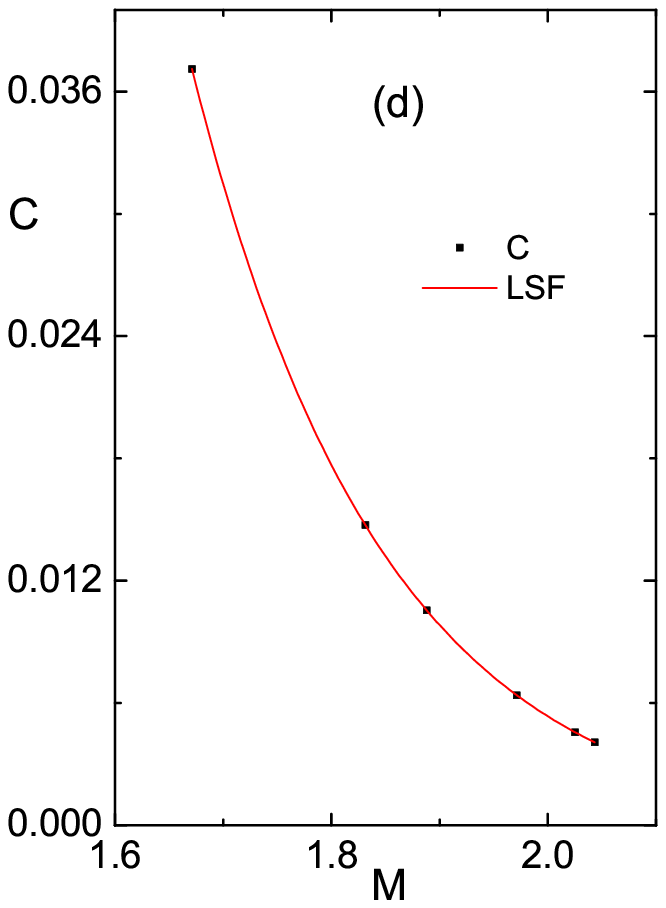}
\caption{(a) $S(M)$, (b) $H(M)$, (c) $D(M)$, and (d) $C(M)$, by
varying $c$ for a fixed radius $R=11.5$ Km (see text,
Eqs.~(\ref{eq-23})-(\ref{eq-25})).} \label{fig:fig4}
\end{figure}

\begin{figure}[h]
\includegraphics[height=7.5cm,width=7.cm]{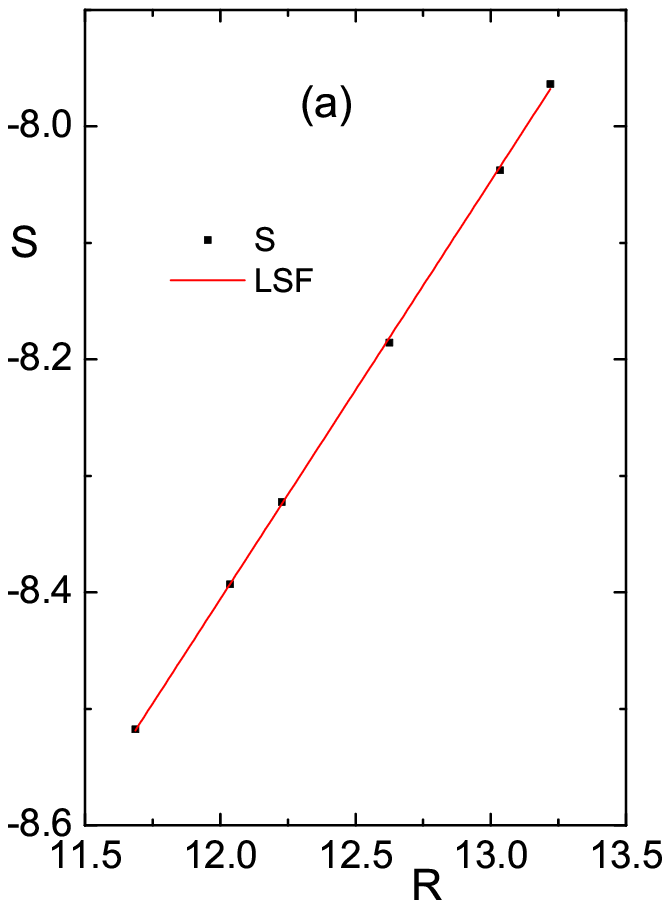}
\includegraphics[height=7.5cm,width=7.cm]{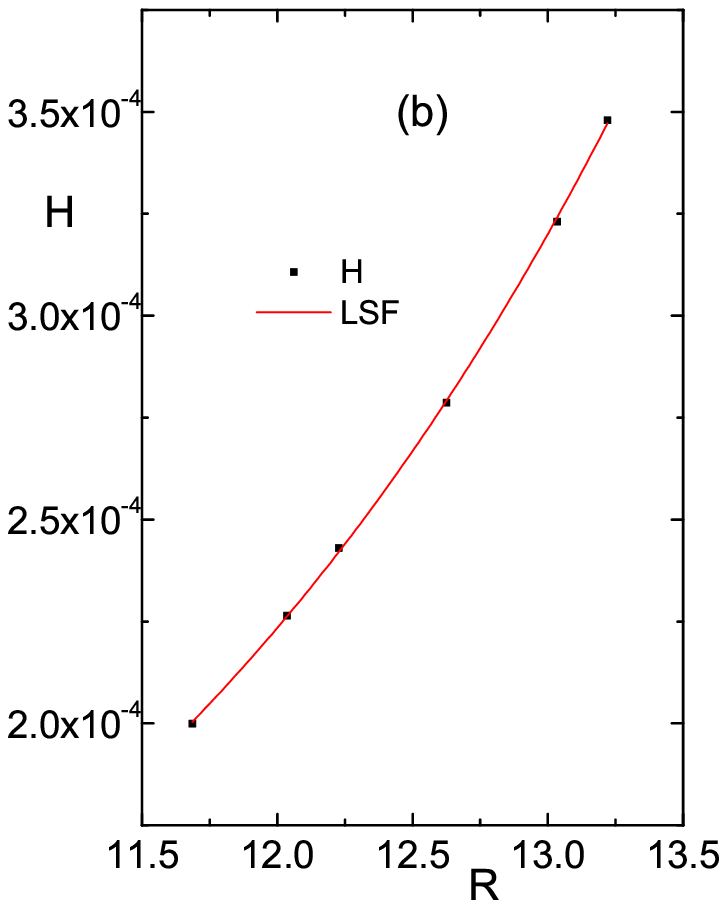}
\\
\includegraphics[height=7.5cm,width=7.cm]{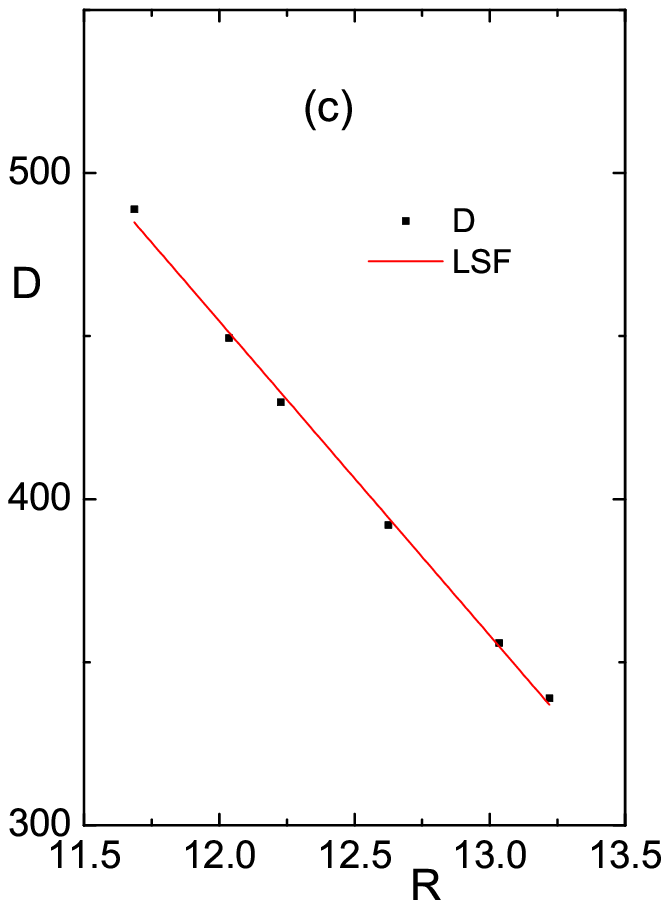}
\includegraphics[height=7.5cm,width=7.cm]{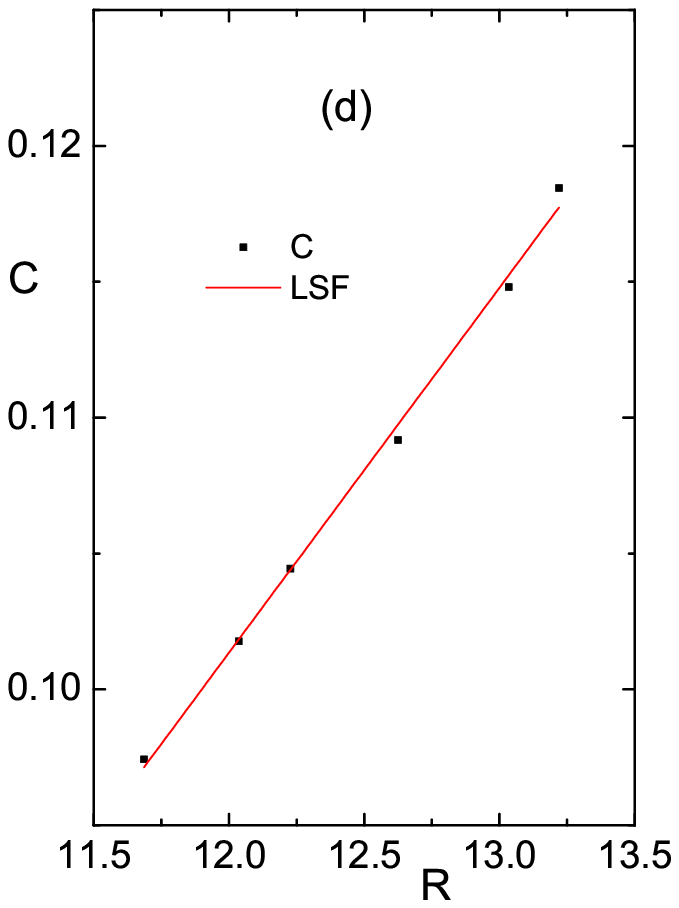}
\caption{(a) $S(R)$, (b) $H(R)$, (c) $D(R)$, and (d) $C(R)$, by
varying $c$ for a fixed mass $M=1.5$ M$_{\odot}$ (see text,
Eqs.~(\ref{eq-26})-(\ref{eq-28})).} \label{fig:fig5}
\end{figure}

In order to study in more detail the connection between the
information and complexity measures with the nuclear interaction
and gravity, we plot $S$, $D$ and $C$ (which correspond to a
maximum mass of a neutron star), as functions of $R$ and $M$,
varying $c$ and $G$ respectively. In Fig.~\ref{fig:fig4} and
Fig.~\ref{fig:fig5} we present the results for the effect of the
nuclear interaction in two cases. This is done by modifying the
equation of state by varying $c$ from 0.7 to 1.5 and then we plot
information and complexity measures, first as functions of $M$ in
Fig.~\ref{fig:fig4} (for a fixed value of $R=11.5$ Km) and second,
of $R$ (for a fixed value of $M=1.5$ M$_{\odot}$) in
Fig.~\ref{fig:fig5}.

Therefore, by keeping $R$ fixed and studying the dependence of $S,
D$ and $C$ on the nuclear interaction indirectly employing $M$, we
note that $S$ is almost a linear decreasing function of $M$
(Fig.~\ref{fig:fig4}(a)), $D$ is increasing exponentially
(Fig.~\ref{fig:fig4}(c)), resulting to a fast exponential decrease
of $C$ versus $M$ (Fig.~\ref{fig:fig4}(d)). Since $R$ is fixed and
so does the volume, increasing $M$ corresponds to a more localized
energy density, hence $S$ decreases with decreasing $D$.

The approximate linear and exponential expressions for $S(M)$,
$D(M)$ and $C(M)$ are (Fig.~\ref{fig:fig4}, $R=11.5$ Km):
\begin{align}
    S &=-7.371 M+2.581, \label{eq-23}\\
    D &=7.027 \, \e^{2.307 M}+313.550, \\
    C &=454.949 \, \e^{-5.624 M}-0.001. \label{eq-25}
\end{align}

Approximate expressions have been obtained by the application of
the least squares fitting (LSF) method. We use linear relations
for fitting of the form $y=c_1 x+c_2$, while exponential relations
are  of the form $y=c_1 \, \e^{-x/c_2}+c_3$.

On the other hand, we keep $M=1.5$ M$_{\odot}$ and we see that all
measures $S$, $D$, and $C$ are linearly depended on $R$, as $c$
varies. Information entropy $S$ increases
(Fig.~\ref{fig:fig5}(a)), while disequilibrium $D$ decreases
(Fig.~\ref{fig:fig5}(c)). For fixed $M$ an increasing $R$
corresponds to a larger volume. Then,  the star becomes more
extended and accordingly, its energy density becomes less
localized. Complexity $C$ increases as a result of that
delocalization, since the system becomes less ordered
(Fig.~\ref{fig:fig5}(d)).

The approximate (fitted) linear expressions for $S(R)$, $D(R)$ and
$C(R)$ are (Fig.~\ref{fig:fig5}, $M=1.5$ M$_{\odot}$):
\begin{align}
    S &=0.359 R-12.713, \label{eq-26}\\
    D &=-96.405 R+1611.450, \\
    C &=0.013 R-0.060. \label{eq-28}
\end{align}

\begin{figure}[h]
\includegraphics[height=7.5cm,width=7.cm]{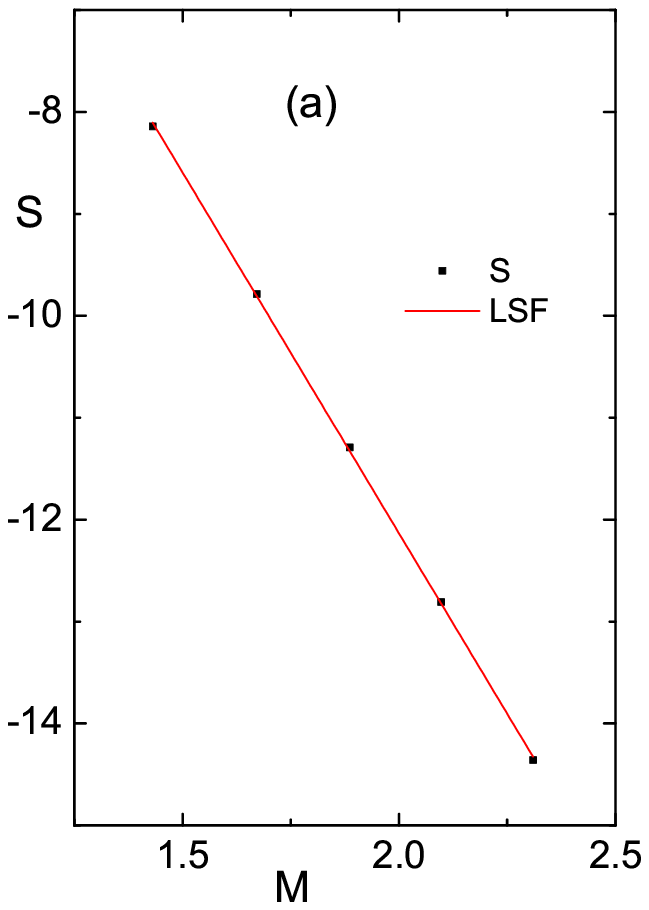}
\includegraphics[height=7.5cm,width=7.cm]{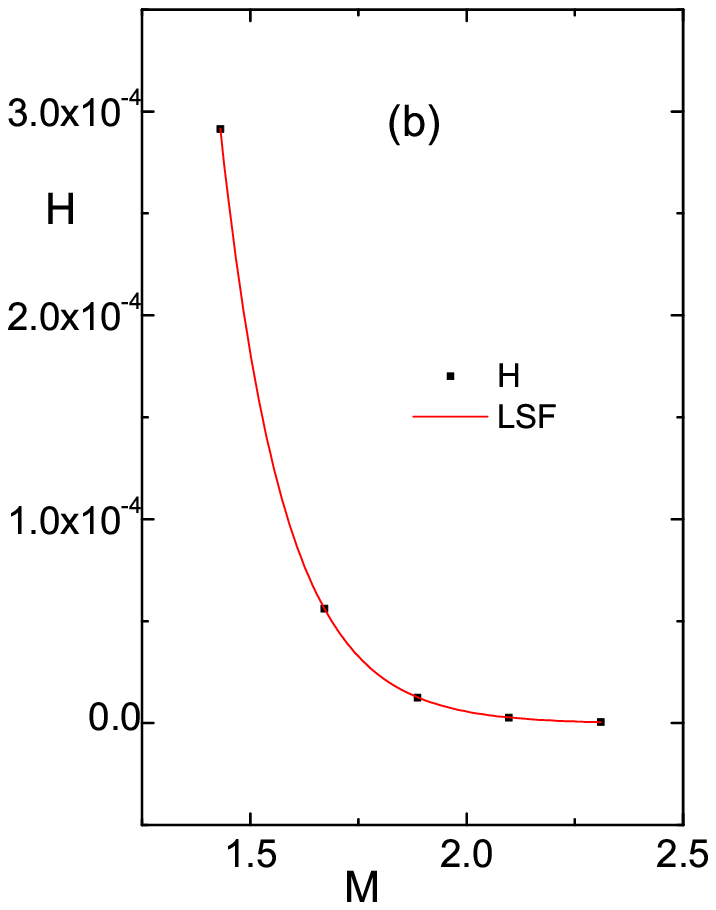}
\\
\includegraphics[height=7.5cm,width=7.cm]{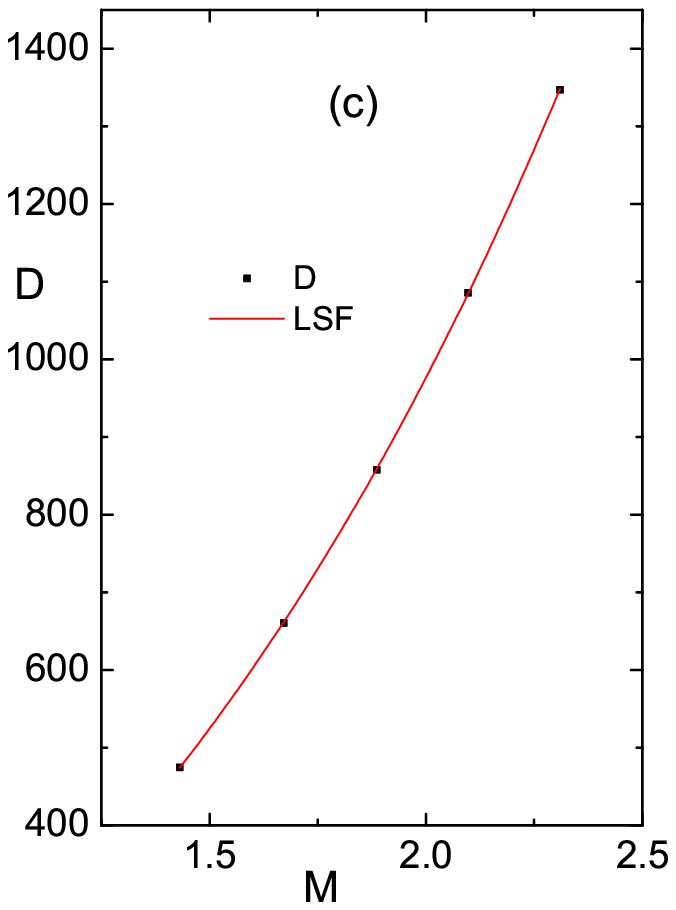}
\includegraphics[height=7.5cm,width=7.cm]{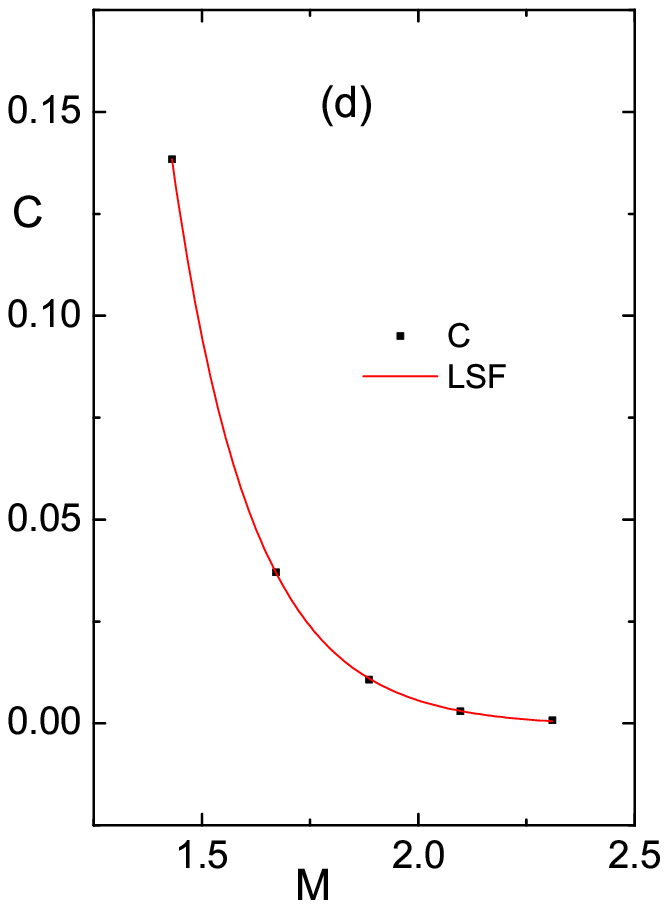}
\caption{(a) $S(M)$, (b) $H(M)$, (c) $D(M)$, and (d) $C(M)$, by
varying $G$ for a fixed radius $R=11.5$ Km (see text,
Eqs.~(\ref{eq-29})-(\ref{eq-31})).} \label{fig:fig6}
\end{figure}

\begin{figure}[h]
\includegraphics[height=7.5cm,width=7.cm]{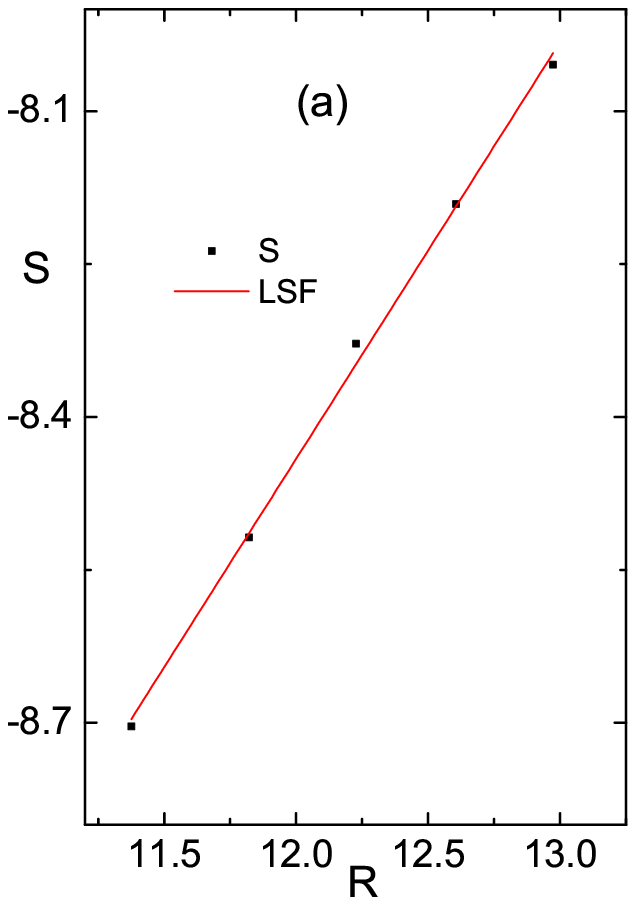}
\includegraphics[height=7.5cm,width=7.cm]{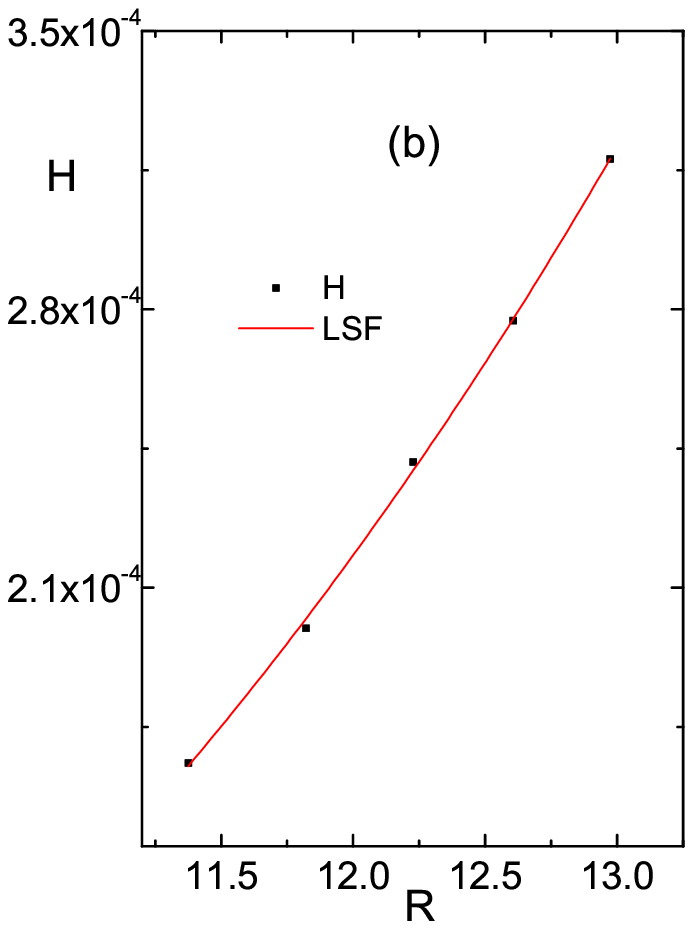}
\\
\includegraphics[height=7.5cm,width=7.cm]{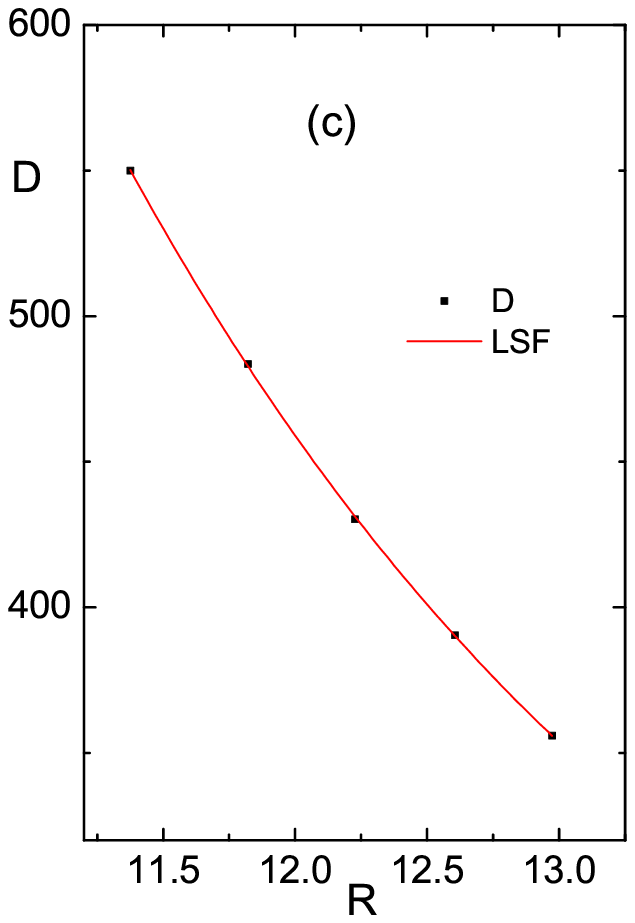}
\includegraphics[height=7.5cm,width=7.cm]{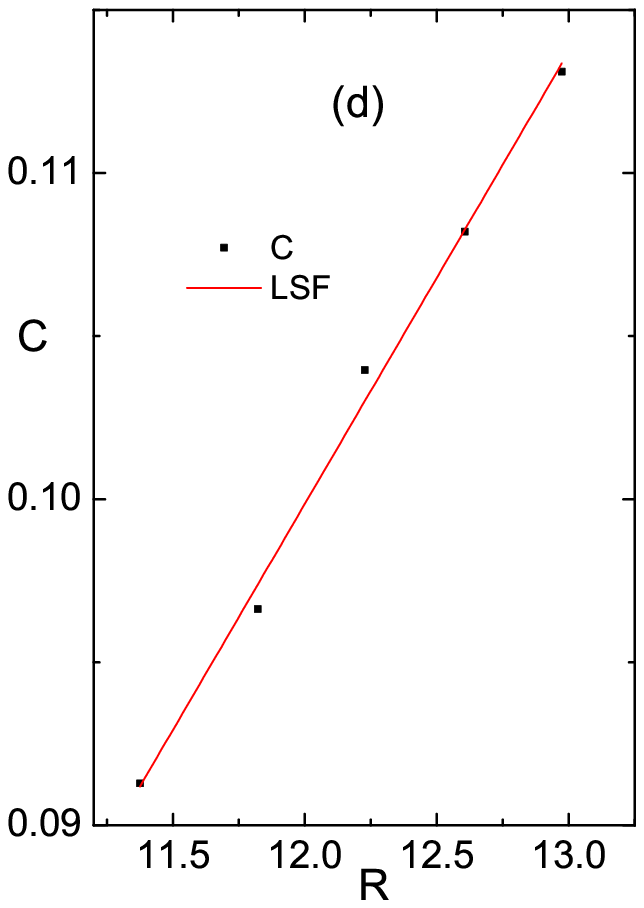}
\caption{(a) $S(R)$, (b) $H(R)$, (c) $D(R)$, and (d) $C(R)$, by
varying $G$ for a fixed mass $M=1.5$ M$_{\odot}$ (see text,
Eqs.~(\ref{eq-32})-(\ref{eq-34})).} \label{fig:fig7}
\end{figure}

We repeat the same series of calculations examining this time the
gravitational dependence of information and complexity measures,
by varying the gravitational parameter $G$ in the range from 0.9
$G$ to 1.1 $G$, while the equation of state is fixed. Our aim is
to see how the variation of $G$ affects quantitatively $S$ in a
neutron star and compare with the results of the previous case.
These results are presented in Fig.~\ref{fig:fig6} and
Fig.~\ref{fig:fig7}. A general comment is that the two cases are
almost equivalent. The trends and the behavior of $S$, $D$ and $C$
obtained by varying the gravitational parameter $G$, keeping  the
equation of state fixed, are almost the same with the
corresponding trends obtained by varying the equation of state for
fixed $G$.

Specifically, first, for fixed $R=11.5$ Km, $S$ decreases linearly
with $M$ (Fig.~\ref{fig:fig6}(a)), $D$ increases exponentially
(Fig.~\ref{fig:fig6}(c)), resulting to a fast exponential decrease
of $C$ with $M$ (Fig.~\ref{fig:fig6}(d)). The energy density of
the system becomes more localized as $G$ increases for a fixed $R$
(fixed volume).

The (fitted) approximate linear and exponential expressions for
$S(M)$, $D(M)$, and $C(M)$ are (Fig.~\ref{fig:fig6}, $R=11.5$ Km):
\begin{align}
    S &= -7.081 M+2.027, \label{eq-29} \\
    D &=362.941 \, \e^{0.712 M}-531.607, \\
    C &=336.821 \, \e^{-5.446 M}. \label{eq-31}
\end{align}

Second, for fixed $M=1.5$ M$_{\odot}$, $S$ increases linearly with
$R$ (Fig.~\ref{fig:fig7}(a)), $D$ decreases exponentially
(Fig.~\ref{fig:fig7}(c)), resulting to a linear increase of $C$
with $R$ (Fig.~\ref{fig:fig7}(d)). The energy density of the
system becomes less localized as $G$ increases for a fixed $M$, as
a result of the radius increase.

The approximate linear and exponential expressions for $S(R)$,
$D(R)$, and $C(R)$ are (Fig.~\ref{fig:fig7}, $M=1.5$ M$_{\odot}$):
\begin{align}
    S &=0.409 R-13.348, \label{eq-32} \\
    D &=41451.805 \, \e^{-0.407 R}-143.591, \\
    C &=0.014 R-0.067. \label{eq-34}
\end{align}

\begin{figure}[h]
\centering
\includegraphics[height=7.5cm,width=7.cm]{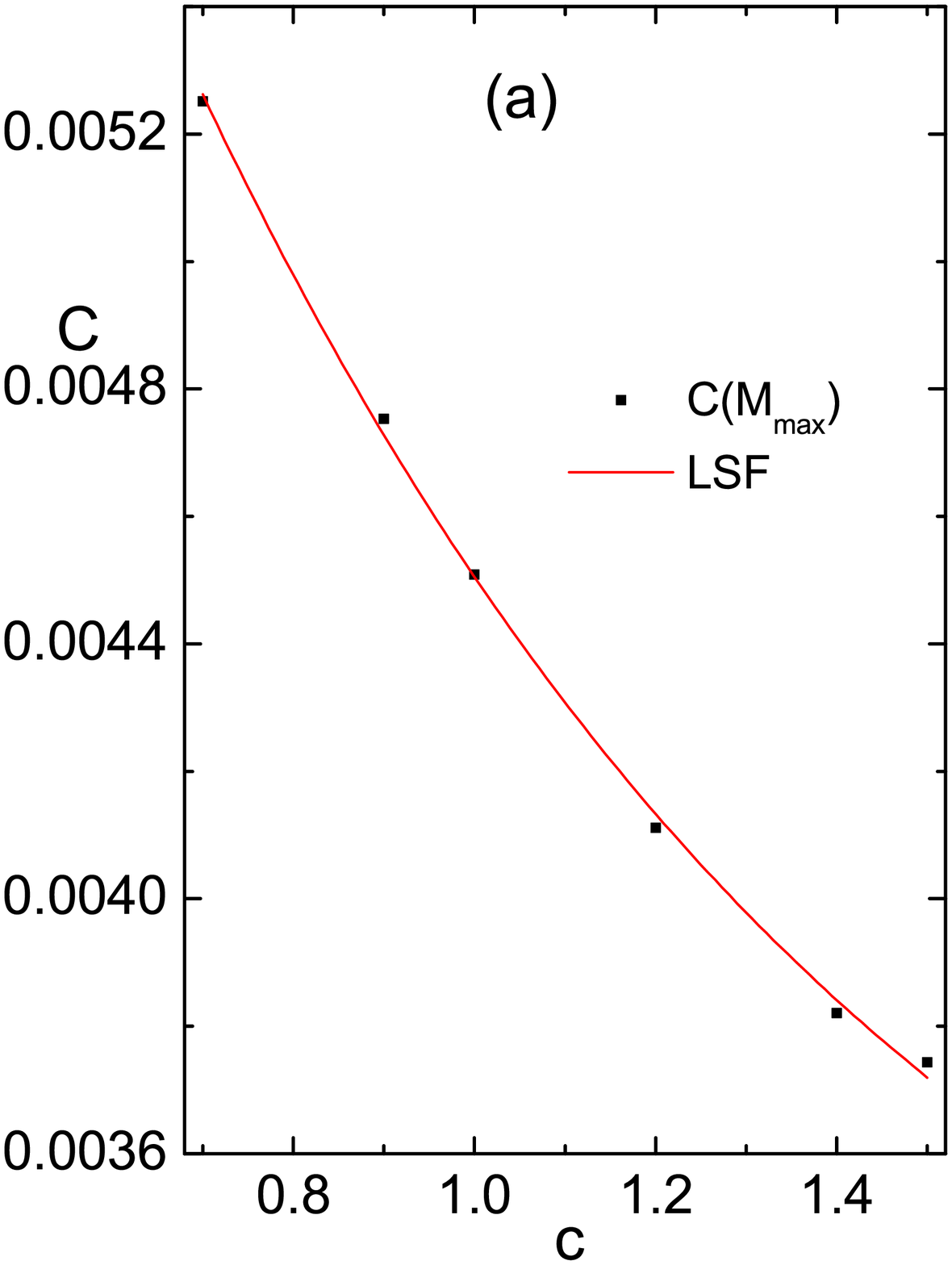}\hspace{-0.8cm}
\includegraphics[height=7.5cm,width=7.cm]{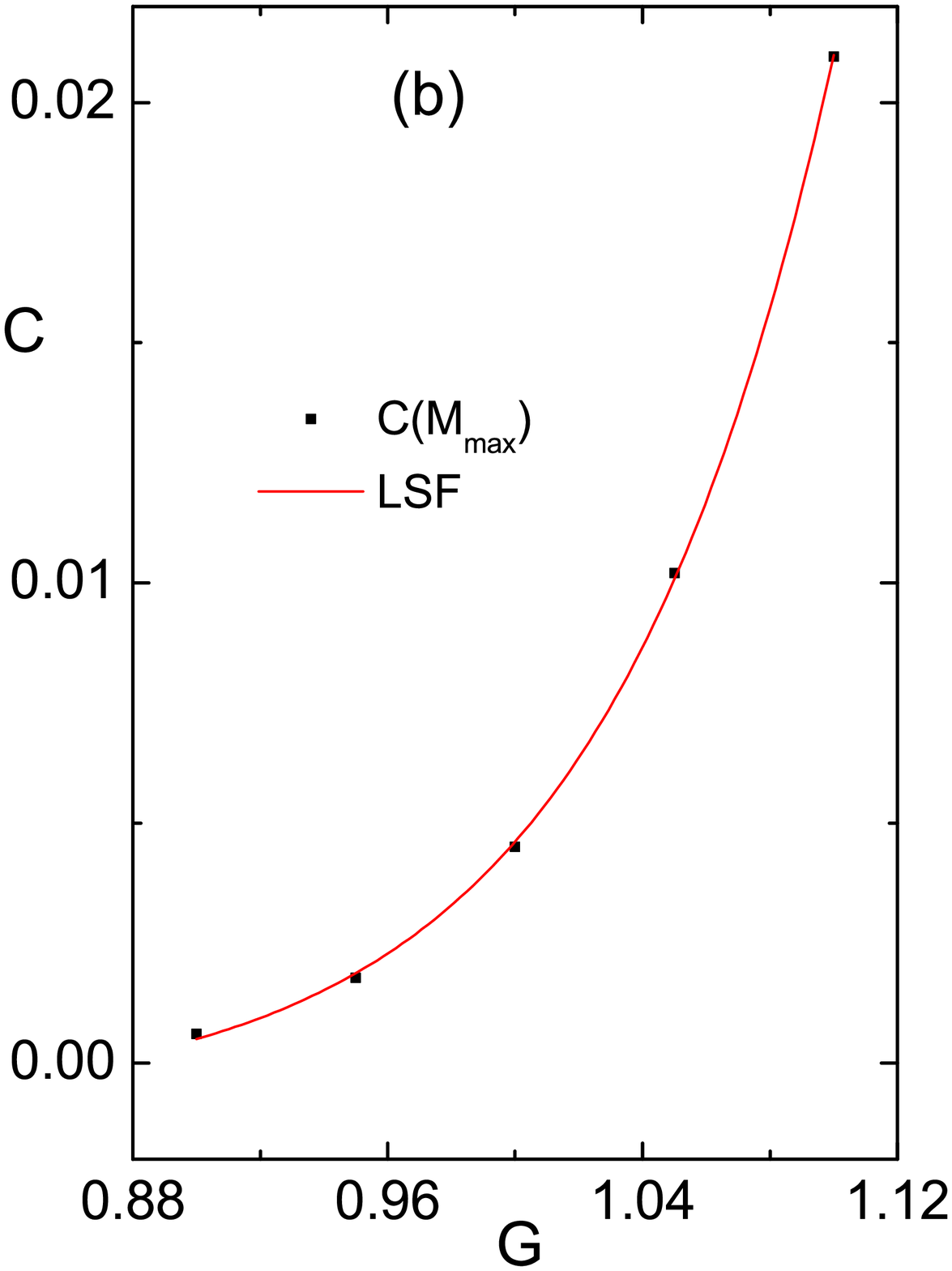}
\caption{(a) Complexity vs the equation of state parameter $c$,
and (b) Complexity vs the gravitational parameter $G$, for a given
$M=M_{\rm max}=1.5$ M$_{\odot}$ (see text,
Eqs.~(\ref{eq-35})-(\ref{eq-36})).} \label{fig:fig8}
\end{figure}

Finally in Fig.~\ref{fig:fig8}, we present the direct dependence
of complexity $C$  on the parameters $c$ and $G$. It is seen from
Fig.~\ref{fig:fig8}(a) that complexity for a given $M_{\rm max}$
is a decreasing function of the equation of state parameter $c$
(the trend is equivalent with the one in Fig.~\ref{fig:fig4}(d)),
while it increases exponentially with the parameter of the
gravitational field Fig.~\ref{fig:fig8}(b).

The corresponding (fitted) exponential expressions for $C(c)$, and
$C(G)$ are:
\begin{align}
    C &=0.006 \, \e^{-1.208 c}+0.003, \label{eq-35} \\
    C &=5.45 \times 10^{-9} \, \e^{13.822 G}-0.001. \label{eq-36}
\end{align}

\section{Summary}\label{sec:4}

We present a study of neutron stars from the point of view of
information and complexity theories. It is shown that the measures
of information entropy $S$ and disequilibrium $D$ can serve as
indices of structure of a neutron star. More specifically, $S$ is
a decreasing function of the mass of the star, while it is an
increasing one of its radius. This result is consistent with the
fact that as a neutron star's mass increases, its radius decreases
resulting to more localized energy and mass densities. The
disequilibrium $D$ shows an inverse behavior. It is an increasing
function of the mass and a decreasing one of its radius. More
localized energy and mass densities correspond to a distribution
far from equiprobability and as a result the disequilibrium of the
system is higher i.e. it is far from equilibrium.

The complexity $C$ of a neutron star is a decreasing function of
its mass. It almost vanishes for a vast set of pairs of
values$(R,M)$,  while it increases rapidly for masses less than
1.5 $M_{\odot}$ and radii greater than 12 Km. But this is a not
such a favorable case for a neutron star compared with
astronomical observations done so far. The favorable one, for
masses larger than 1.5 $M_{\odot}$ and radii less than 12$\,$ Km
corresponds to almost vanishing complexity, supporting the
conclusion that a neutron star is an ordered system, which cannot
grow in complexity as its mass increases.

Furthermore, we investigate the impact of the equation of state
parameter $c$ and the gravitational parameter $G$ on  $S$ and
 $C$. The behaviors of information and complexity
measures are equivalent in both cases. Complexity decreases
exponentially with the mass, while it increases linearly with the
radius. In direct calculations, complexity decreases exponentially
with the equation of state parameter $c$, while it increases
exponentially with the gravitational parameter $G$.

\section*{Acknowledgements}

K. Ch. Chatzisavvas is supported by a Post-Doctoral Research
Fellowship of the Hellenic State Institute of Scholarships (IKY).

 \clearpage
\newpage

\section{Figures}

\clearpage
\newpage

\clearpage
\newpage

\clearpage
\newpage

\clearpage
\newpage

\end{document}